\documentclass[
 reprint,
 amsmath,amssymb,
 aps,
]{revtex4-2}
\usepackage[utf8]{inputenc}
\usepackage{graphicx}
\usepackage{dcolumn}
\usepackage{orcidlink}
\usepackage{subfigure}
\usepackage[title]{appendix}
\usepackage{bm}
\usepackage{xcolor}
\usepackage{hyperref}
\begin{document}

\preprint{APS/123-QED}

\title{Maximal hypersurface in a D-dimensional dynamical spacetime}

\author{Suraj Maurya\orcidlink{0000-0001-6907-8584}}
 \email{p20200471@hyderabad.bits-pilani.ac.in (S. Maurya)}
 
 \author{Rahul Nigam\orcidlink{0000-0002-0497-5898}}
 \email{rahul.nigam@hyderabad.bits-pilani.ac.in (R. Nigam)}

 \author{Sashideep Gutti\orcidlink{0000-0001-7555-8453}}
 \email{sashideep@hyderabad.bits-pilani.ac.in (S. Gutti)}
\affiliation{Birla Institute of Technology and Science Pilani (Hyderabad Campus), Hyderabad 500078, India}

\date{\today}

\begin{abstract}
  In this article, we set up a variational problem to arrive at the equation of the maximal hypersurface in the interior of a spherically symmetric evolving trapped region. In the first part of the article, we present the Lagrangian and the corresponding Euler-Lagrange equations that maximize the interior volume of a trapped region that is formed dynamically due to infalling matter in D-dimensions, with and without the cosmological constant.  In the second part, we explore the properties of special radii, which we call Reinhart radii, that play a crucial role in approximating the maximal interior volume of a black hole. We derive a formula to locate these Reinhart radii in terms of coordinate invariants like area radius, principle values of the energy-momentum tensor, Misner-Sharp mass, and cosmological constant. Based on this formula, we estimate the location of Reinhart radii in various scenarios: (a)  the case of static BTZ black holes in $(2+1)$-dimensions and for the Schwarzschild,  Schwarzschild-de Sitter, and Schwarzschild-anti-de Sitter black holes in D-dimensions. We plot the location of the Reinhart radii in relation to the event horizon and cosmological horizon in a static D-dimensional scenario, (b) cosmological case:  we prove that these Reinhart radii do not exist for homogeneous evolving dust for the zero and negative cosmological constant but exist in the presence of positive cosmological constant when the scale factor is greater than a critical value. We also show the relation between these Reinhart radii and Kodama vectors. 
\end{abstract}

\maketitle

\section{Introduction}  
One of the most intriguing and mysterious objects in the Universe is the black hole. This area of research is now at the forefront ever since the black holes have been observationally identified using gravitational wave astronomy. The study of black holes is now considered mainstream and astrophysically relevant. The thermodynamical properties of the black holes, the singularities inside the black holes, etc. are yet to be understood fully. The information loss of the black hole and its resolution are still sought after mysteries.\par
One of the intriguing aspects of a black hole is the volume in the interior of a black hole. This question has been addressed by various authors using different approaches. Maulikh Parikh \cite{MP} discusses the definition of volume by constructing an invariant slice of the spacetime inside the black hole horizon. Cvetic \textit{et al}. \cite{Gibbons} have discussed the thermodynamical volume, $V_{th}$ inside a black hole in the presence of a varying cosmological constant $\Lambda$. $V_{th}$ is defined as the conjugate variable to $\Lambda$ appearing in the first law of thermodynamics for black hole i.e. $dE = TdS + \Omega dJ+\Phi dQ+V_{th}\Lambda$, where $E$ is the gravitational enthalpy of spacetime. Christodoulou and Rovelli \cite{CR} provided a somewhat different definition of the black hole volume, in which the volume grows indefinitely as a function of the advance time.\par
In this article, we generalize the approach due to Christodoulou and Rovelli \cite{CR} for finding the interior volume of a black hole in the case of dynamical situations. They find out the volume inside the black hole via a variational approach. They define a spacelike curve from the event horizon to the singularity, in general, and from among them define an extremal curve or a maximal hypersurface that yields the maximum volume in the interior of the black hole thus setting up a Lagrangian framework to solve the above question. A similar method is adopted in the following papers \cite{Bibhas, BZ, BJ, Ong, YCO, CL, XY, MZ, SSR} discussed by various authors.\par 
In this article, we work on two aspects concerning the volume of black holes.  The first part deals with the question about the evolution of the interior volume of the trapped region where the black hole is in the process of formation. Due to the evolution of the trapped region, the interior volume of the black hole evolves.  We set up a variational problem where we write down a Lagrangian for the spacelike curve between the apparent horizon and the singularity, using which we can estimate the maximal volume. We then obtain the equations for this maximal hypersurface from the Euler-Lagrange equations for the obtained Lagrangian. We first solve the problem for the simple case of $(2+1)$-dimensions where the underlying equations and analysis are simpler. We then carry out the analysis for the case of D-dimensional spherically symmetric dust evolution that leads to the formation of the black hole.\par
It is proved in \cite{CR} that the volume generated by maximal hypersurface has a maximum contribution from a certain region that we call in this article Reinhart radius which we denote as $R_{R}$ (the subscript stands for ``Reinhart").  This region provides an excellent approximation for the interior volume of the black hole \cite{CR}. For a Schwarzschild black hole with mass $M$, the event horizon is at $R=2M$ while the value of $R_{R}=3M/2$. This Reinhart radius $R_{R}$ lies in the interior of the black hole. This region inside the Schwarzschild black hole which is also a maximal hypersurface was first discovered by Reinhart \cite{Reinhart} in 1973. Similar points of interest in the interior of a black hole have been found in the examples described below in various other black holes like BTZ, Kerr, and Kerr-anti-de Sitter black holes \cite{SSR, BJ, XY}. The volume of a Schwarzschild black hole is shown to be equal to $V=3\sqrt{3}M^2v$ where $v$ is the advance time \cite{CR}.  The special feature of the hypersurface is that the normal to the surface is divergence-free, i.e., the trace of the extrinsic curvature vanishes. The Reinhart radii have played a pivotal role in the approximation of the volume of black holes in a more general setting. For instance, the asymptotic volume of a static BTZ black hole is crucially dependent on a point $R_{R}=\sqrt{M/2|\Lambda|}$ discussed in \cite{MZ} with $M$ being the ADM mass of the BTZ black hole and $\Lambda$  is the cosmological constant. Even in the presence of rotation, the asymptotic formula for the interior volume of the BTZ black hole crucially depends on the point $R_R=\sqrt{M/2|\Lambda|}$ as shown in  \cite{SSR}. For the BTZ black hole with rotation, we found that the maximal interior volume is $V_\Sigma = \pi v \sqrt{M^2/|\Lambda| - J^2}$, where $v$ is advanced time, $M$ is mass and $J$ is the angular momentum of the BTZ black hole. \par
The paper \cite{CL} tackles the problem of time-dependent metrics. The important result from \cite{CL} is the estimation of volume during the evolution due to Hawking evaporation. They prove that the volume follows a monotonically increasing trend in spite of Hawking radiation (till the Planck regime is reached). The volume is given by $ V(v)\approx 3\sqrt{3}\pi m^2_0 v (1- 9B/2m^2_0)$, where $m_0$ is the original mass of the black hole and $B$ is a constant.  Though the result is counterintuitive, the result in \cite{CL} can be understood owing to the presence of Reinhart radius that decreases with time. The volume therefore continues to increase since it is proportional to the advanced time $v$. Though we did not include the evaporating case in our article, we show that our results match in all situations where  Reinhart radius exists. \par
When one explores the interesting aspects of black hole interiors one thinks of the interior as a somewhat trivial region with the only interesting feature being the spacetime singularity (and the inner horizon in case of rotating or charged black holes). The presence of $R_{R}$ between the event horizon and singularity, therefore reveals yet another interesting region in the interior of the black hole. These regions have not been explored in all its generality.  In this article, we explore the features of the Reinhart radii in various spacetimes. We point out the relation of the Reinhart radius with the Kodama vector. We show that these radii correspond to the locations in the maximal hypersurface that are tangential to the Kodama vector.  The study of Reinhart radii is exhaustive where we explore its evolutionary aspects in D-dimensions with and without cosmological constant. We arrive at interesting results for the various cases discussed. We develop a formula for tracking the evolution of the $R_{R}$ from which we can deduce its location.\par
In Sec. II we review the features of spacetime in D-dimensions and the evolution of dust. In Sec. III $A$, we review the evolving dust model in $(2+1)$-dimensions. In Sec. III, we set up the Lagrangian formulation to locate the maximal hypersurface that maximizes volume for the homogeneous dust model for $(2+1)$-dimensional and D-dimensional cases. In Sec. IV, we use the extrinsic curvature method to estimate the Reinhart radius in a $(2+1)$-dimensional case. In the subsections of Sec. IV, we discuss the vacuum case, static black hole case, and cosmological case in $(2+1)$-dimensions. In Sec. V, we discuss the extrinsic curvature method to locate the Reinhart radius in D-dimensions. In the subsections of Sec. V, we discuss the vacuum scenario, the D-dimensional Schwarzschild black hole case, the D-dimensional Schwarzschild deSitter and anti-deSitter case, and the cosmological case. In Sec. VI we discuss the estimation of the volume of evolving black hole in D-dimensions. In Sec. VII  we discuss the relation between the Reinhart radii and Kodama vector. In Sec. VIII, we discuss the conclusions of the work. We discuss the solution of the scale parameter $a(t)$ for the homogeneous dust evolution in the Appendix.
\section{D-dimensional evolving dust scenario}
In this section, we review the evolving dust model in D-dimensional spherically symmetric spacetime. The discussion is inclusive of a cosmological constant. The general metric for a D$( = n+2)$-dimensional spherically symmetric spacetime is of the form
    \begin{equation}
    ds^2 = - e^{\mu(t,r)}dt^2 + e^{\lambda(t,r)}dr^2 + R^2 (t,r)~d{\Omega}_{n}^{2}
    \end{equation}
where $d{\Omega}^2_{n} = d{\theta}^2_{1} + sin^2{\theta}_{1}(d{\theta}^2_{2} +sin^2{\theta}_{2} (d{\theta}^2_{3} + ... + sin^2{\theta}_{n-1}d{\theta}^2_{n}))$ is the metric on unit $n$-dimensional sphere, $t$ is the time coordinate and $r$ is the comoving radial coordinate. It is easily shown in \cite{Rakesh} that the $g_{00}$ component of the metric can be chosen to be minus one, i.e., $g_{00} = -1$ when the matter considered is dust. The metric then has a simpler form given by
    \begin{equation}\label{metric}
    ds^2 = -dt^2 + e^{\lambda(t,r)}dr^2 + R^2 (t,r)~d{\Omega_n}^2
    \end{equation}
The Einstein field equations are given below, 
    \begin{equation}
    G_{\mu\nu} + \Lambda g_{\mu\nu} = R_{\mu\nu} - \frac{1}{2}Rg_{\mu\nu} + \Lambda g_{\mu\nu} = \kappa T_{\mu\nu}
    \end{equation}
Here, $G_{\mu\nu}$ is the Einstein tensor, $R_{\mu\nu}$ is the Ricci curvature tensor, $R$ is the Ricci scalar curvature, $g_{\mu\nu}$ is the metric tensor, $T_{\mu\nu}$ is the stress-energy tensor and $\kappa$ is the Einstein gravitational constant and is related to Newton's gravitational constant $G_n$ as ($\kappa$ = 8$\pi G_n/c^4$). The matter we are considering here is a pressureless dust, hence, the only nonzero component of the stress-energy tensor (in the comoving and synchronous coordinate system) is $T_{00} = \epsilon (t,r)$, where $\epsilon (t,r)$ is the energy density of the dust. With these conditions we get the Einstein equations which are shown in \cite{Rakesh}. These are listed below
    \begin{multline}
     G_{00} = \frac{e^{-\lambda}}{R^2} \bigg[\frac{n(n-1)}{2} [e^{\lambda} (1+\dot{R}^2) - R'^{2}] +  \frac{n}{2} R R' \lambda'\\ -\Lambda e^{\lambda} R^{2} +  \frac{n}{2} (-2R R'' + e^{\lambda} R \dot{R} \dot{\lambda}) \bigg] = k \epsilon(t,r)
    \end{multline}
    \begin{equation}
    G_{01} = \frac{n}{2}\frac{(R' \dot{\lambda} - 2 \dot{R}')}{R} = 0
    \end{equation}
    \begin{multline}
    G_{11}  = \frac{1}{R^{2}} \bigg[\frac{n(n-1)}{2} (R'^{2} - e^{\lambda}(1+\dot{R}^2))\\ + \Lambda e^{\lambda} R^{2} - n e^{\lambda} R \ddot{R}  \bigg] = 0
    \end{multline}
    \begin{multline}
    G_{22} = - \frac{1}{4} e^{-\lambda} \bigg[2(n-2)(n-1) [e^{\lambda}(1+\dot{R}^2) - R'^{2}]\\ - 2(n-1)[2RR'' - RR'\lambda'- e^{\lambda}(R\dot{R}\dot{\lambda} + 2R\ddot{R})]\\ + e^{\lambda}R^{2}(-4\Lambda + \dot{\lambda}^2 + 2\ddot{\lambda}) \bigg] = 0. 
    \end{multline}
The other nonzero relations are given by 
    \begin{equation}
    G_{(j+1~j+1)} = sin^2{\theta_{(j-1)}} G_{(jj)},
    \end{equation}
where $j$ takes values from 2 to $n+1$. The expressions for the evolution of matter can be obtained by simplifying the above set of equations. Solving for the $G_{01}$, we get
    \begin{equation} \label{elambda}
    e^{\lambda} = \frac{R'^{2}}{1 + f(r)},
    \end{equation}
where the integration constant $f(r)$ is an arbitrary function called the energy function. Integration of the $G_{11}$, equation after using the above relation, gives
    \begin{equation}\label{rdotsquare}
    \dot{R}^2 = f(r) + \frac{2\Lambda}{n(n+1)} R^2 + \frac{F(r)}{R^{(n-1)}},
    \end{equation}
where $F(r)$ is called the mass function. Solving for $G_{00}$, we find 
    \begin{equation}\label{density}
    \kappa \epsilon(t,r) = \frac{n F'}{2 R^n R'}.
    \end{equation}
This gives us the expression for the mass function  as
    \begin{equation}\label{massfunction}
    F(r) = \frac{2\kappa}{n} \int \epsilon(0,r) r^n dr,
    \end{equation}
where $\epsilon(0,r)$ is the initial energy density of the dust.We make a choice that when the comoving time $t=0$, we set the comoving radius equal to the area radius, $R=r$. We work for the case of marginally bounded shells of dust where we require that $f(r) = 0$. The result (\ref{massfunction}) is obtained by keeping the constant value of $f(r) = 0$, and this holds true from here on. Throughout the article, we shall assume that $\epsilon >0$. (weak energy condition is satisfied).\\
We consider the scenario where the dust cloud is of finite extent. We denote the outermost comoving label to be $r_0$. The region exterior to radius $r_0$ is the vacuum. The metric element in the exterior of the D($=n+2$)-dimensional black hole with a cosmological constant  given by \cite{Rakesh}
\begin{multline}
    ds^2 = -\bigg(1-\frac{F(r_0)}{R^{n-1}}-\frac{2\Lambda R^2}{n(n+1)}\bigg)dT^2\\ + \bigg(1-\frac{F(r_0)}{R^{n-1}}-\frac{2\Lambda R^2}{n(n+1)}\bigg)^{-1}dR^2 + R^2d\Omega^2_n,
    \label{exteriormetric}
\end{multline}
where, $F(r_0)$ is the mass function evaluated at $r_0$. $T$ is like the Schwarzschild time coordinate and $R$ is the area radius.  The above metric is obtained by matching the interior with the exterior metric across the boundary $r_0$.  For $n=2$ i.e., the Schwarzschild black hole, $F(r_0)=2GM$, where $M$ is the ADM mass. In general D-dimensions, the relation between mass function $F(r)$ and mass of the dust cloud $M$ in $(n+1)$ spatial dimensions \cite{Rakesh} is defined as
\begin{equation}\label{9}
    F(r) = \frac{2\kappa}{n}\frac{M\Gamma(\frac{n+1}{2})}{2\pi^{\frac{n+1}{2}}},
\end{equation}
where $\kappa = 8\pi G/c^4$ is the Einstein constant and $G$ is the Newton's gravitational constant.
Now we track the evolution of the maximal volume in the next section.

\section{Lagrangian formulation for the maximal volume of an evolving black hole}
In this section, we derive the Lagrangian and the corresponding Euler-Lagrange equation that maximizes the volume inside an evolving black hole.  We first discuss the evolving dust models in $(2+1)$-dimensions. The reason for separating the $(2+1)$-dimensional case from the general D-dimensional case is that there are unique features in the $(2+1)$-dimensional case that do not generalize to general D-dimensions. 
\subsection{Review of (2 + 1)-dimensional evolving dust solution}
In this subsection, we study a dynamical situation corresponding to the formation of the BTZ black hole.  We set the angular momentum of the BTZ black hole to zero in order to obtain analytically tractable expressions since we do not yet have an analytical collapsing model that yields a rotating BTZ black hole. 
The Einstein equations for the $(2+1)$-dimensions can be explicitly solved. The black hole in $(2+1)$-dimensions occurs only in the case of the negative cosmological constant. The solution given in \cite{rossandmann, sashideep} is given by
\begin{multline}\label{metric1}
ds^2=-dt^2+\frac{(\cos(\sqrt{|\Lambda| }t)+B'\sin(\sqrt{|\Lambda| }t))^2dr^2}{|\Lambda| r^2+|\Lambda| B^2-2\kappa\int_{0}^{r}{\epsilon _i(s) sds}+1}\\ +(r\cos(\sqrt{|\Lambda| }t)+B\sin(\sqrt{|\Lambda| }t))^2d\phi^2,
\end{multline}
 As the cosmological constant $\Lambda=-1/l^2$ is negative, the background space is AdS. The metric is expressed in terms of two functions of comoving $r$, \cite{sashideep}, $B(r)$ and $\epsilon_i(r)$. Here $R(r,t) = r\cos(\sqrt{|\Lambda| }t)+B\sin(\sqrt{|\Lambda| }t)$  is the area radius defined geometrically using the Killing vector $\partial/\partial \phi$ such that the perimeter of the shell of comoving shell $r$ is $2 \pi R$. $B(r)$ decides the initial velocity of the dust cloud and $\epsilon_i(r)$ decides the initial density. As the cloud evolves, the density evolves and the area radius of each shell evolves (decreases if the cloud undergoes a collapse scenario that leads to the formation of a black hole).  In our present context, we focus on homogeneous dust interior since the model offers simpler equations without compromising the caveats involved. \par
We choose the boundary of the homogeneous dust to be at a comoving coordinate $r=r_0$. Outside this is comoving radius $r_0$; we assume that there is no more dust and hence is the vacuum. We further choose the condition that $B(r)=0$ and $\epsilon_i=|\Lambda|/\kappa$ so that the metric (\ref{metric1}) is in its simplest form given by
\begin{equation}
ds^2=-dt^2+\cos^2(\sqrt{|\Lambda| }t)dr^2+r^2\cos^2(\sqrt{|\Lambda| }t)d\phi^2.
\label{nmetric1}
\end{equation}
We note that $F(r)$ is given by the expression $F(r)=2\kappa\int_{0}^{r}{\epsilon _i(s) sds}$ in $(2+1)$-dimensions. For the parameters that we consider here, it is equal to $|\Lambda|r^2$.  It is easily shown that the metric (\ref{metric1}) can be smoothly matched at the hypersurface $r=r_0$ by equating the first and second fundamental form to the exterior BTZ metric [from Eq. \ref{exteriormetric}], given by
\begin{equation}
ds^2 = -(|\Lambda| R^2 - M) dT^2 + \frac{dR^2}{(|\Lambda| R^2 - M)}  + R^2 d\phi^2,
\label{exterior}
\end{equation}
with $T$ as the time coordinate corresponding to Killing time. Here $M$ is the ADM mass. We note that for the model under consideration that $M= \kappa \epsilon_i r_0^2-1=|\Lambda|r_0^2-1$ \cite{sashideep}, with $r_0$ the outer comoving radius of the dust cloud. It is more convenient to switch from Schwarzschild-like coordinate system $(t,R,\phi)$ to Eddington-Finkelstein coordinates $(v,R,\phi)$ to avoid the coordinate singularity at the horizons. The Eddington-Finkelstein coordinates are defined as
\begin{equation}\label{17}
    v = t + \int^{R}\frac{dR}{N^2(R)}, \ \ R = R \ and \ \phi=\phi,
\end{equation}
where $N^2(R) = (|\Lambda| R^2-M)$. The metric (\ref{exterior}) can now be written as
\begin{equation}
    ds^2 = -(|\Lambda| R^2-M)dv^2 + 2dvdR + R^2d\phi^2,
\end{equation}
\subsubsection{Apparent Horizon:} Our goal in this article is to find out the volume of an evolving black hole. To track the boundary of the evolving black hole, we need the location of the apparent horizon.  As given in \cite{rossandmann, sashideep}, the condition for the expansion parameter for outgoing null geodesics to become zero for a general metric (\ref{metric1}) is given by \cite{sashideep}. So
\begin{equation} 
\frac{2\kappa\int_{0}^{r}{\epsilon _i(s) sds}-1}{|\Lambda| R^2}=1.
\label{trappingcondition1}
\end{equation}
Physically this means that the apparent horizon occurs when the Misner-Sharp mass of a given shell gets compressed to a small area radius given by the above equation. For  the specific case we are considering in this article, we have
\begin{equation} 
\frac{|\Lambda| r^2-1}{|\Lambda| r^2cos^2(\sqrt{|\Lambda|} t)}>1.
\label{trappingcondition2}
\end{equation}
This implies the curve of the apparent horizon is given by the relation
\begin{equation}
r_a^2=\frac{1}{|\Lambda| sin^2(\sqrt{|\Lambda|} t)}.
\label{eqn9}
\end{equation}
We consider the collapsing regime where $t$ goes from $0$ to $\pi/(2\sqrt{|\Lambda|})$. At time $t=0$, we have the apparent horizon at $r_a=\infty$. This is only true for the cosmological case in which we take $r_0$ to infinity.  We note that in the time interval considered during the collapsing phase, the apparent horizon $r_a$ is a decreasing function of time. This implies the shell $r_0$ is trapped first, and then the smaller values of $r$ get trapped.  
\begin{figure}[htp]
    \centering
    \includegraphics[width=0.48\textwidth]{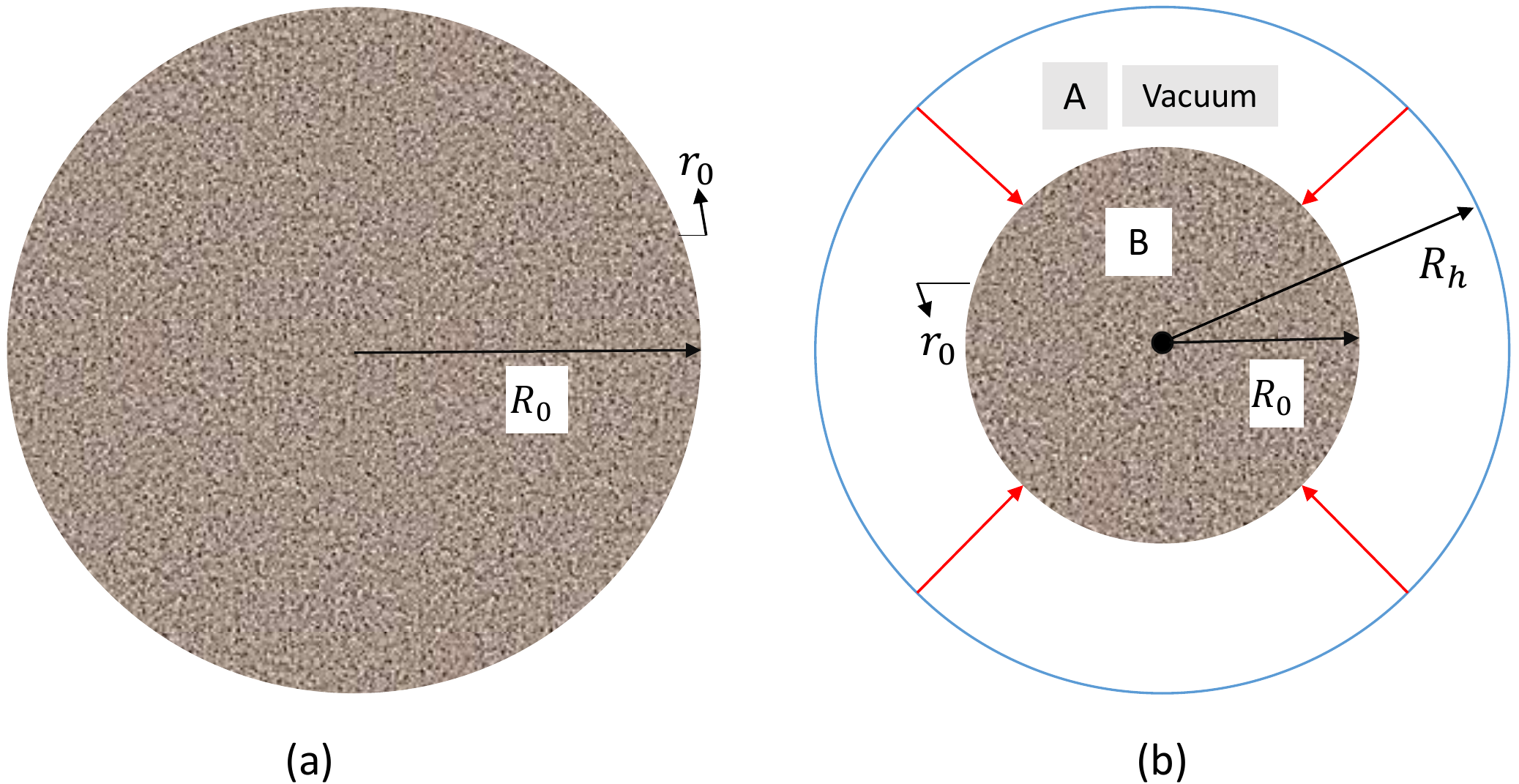}
    \caption{Figure 1(a) shows the homogeneous dust cloud whose outer boundary labeled with the comoving radial coordinate is formed at, $r = r_0$, and Fig. 1(b) shows that this homogeneous dust cloud collapses to form a non-rotating BTZ black hole. Here, $A$ represents the vacuum region that formed during the dust collapse, has a radius of the outer boundary $R_h$, and is known as the event horizon and $B$ represents the evolving dust that eventually collapses to a singularity.}
    \label{fig:1}
    \end{figure}
This also implies that since there is no more mass collapsing beyond the shell $r_0$, the radius $R$ at which $r_0$ becomes trapped is also the event horizon.  The physical radius of the event horizon is given by, $R_h=\sqrt{M/|\Lambda|}$ where $M=|\Lambda| r_0^2-1$. During the course of evolution, we therefore have two distinct regions inside the BTZ black hole as illustrated in Fig. \ref{fig:1}. $(A)$ is the region from the event horizon $R=R_h$ to the area radius of the outer shell $R_0=r_0 cos (\sqrt{|\Lambda|} t)$ of the dust cloud inside the event horizon $R_h$. Region $(B)$ consists of the interior region of the dust cloud. We note that the region $(A)$ is a vacuum whereas $(B)$ is evolving dust that eventually becomes singular. The plan is to track the evolution of the maximal volume where we take into account both the regions $(A)$ and $(B)$. We discuss the evolution of maximal volume in the next subsection. 
\subsection{Lagrangian formulation for the maximal volume of a black hole in (2 + 1)-dimensions}
In this subsection we use the method given in \cite{CR} to construct a two-dimensional spacelike surface $\Sigma$, which is a direct product of one-sphere parameterized by $\phi$ and a curve $\gamma$ embedded in the two-dimensional space parameterized by $\lambda$ in the submanifold  $(v,R)$. We note that this hypersurface extends across the exterior (vacuum spacetime) and the interior (dust cloud). It is more convenient to describe this hypersurface in terms of $(v,R)$ in the exterior spacetime and in coordinates $(t,r)$ in the interior of the dust cloud. In Fig.(\ref{fig:1}), this hypersurface extends from the event horizon in region $(A)$ to the point $R=0$ in region $(B)$. We note the following relations
\begin{equation}\label{eqn10}
    \Sigma \sim \gamma\times S^1, \ \ \gamma\sim [v(\lambda), R(\lambda)] \ and \  \ \gamma\sim [t(\lambda), r(\lambda)]
\end{equation}
We choose $\lambda =0$ at the event horizon $(R_h)$ and call $\lambda_f$ ($f$ for ``final") the value of $\lambda$ at $R=0$. Thus, the initial and final end points of $\gamma$ are given by
\begin{equation}\label{eqn11}
    \begin{split}
     R(\lambda =0) = R_h, \ \  \ \  R(\lambda = \lambda_0) = R_0\\
     v(\lambda = 0)= v, \ \ \ \  \  v(\lambda = \lambda_0) = v_0\\
     t(\lambda=\lambda_0)=t, \ \ \ \  r(\lambda=\lambda_0)=r_0,\\
     t(\lambda=\lambda_f)= \pi/2\sqrt{|\Lambda|}, \ \
     r(\lambda=\lambda_f)= 0. 
          \end{split}
\end{equation}
The surface $\Sigma$ is coordinated by $\lambda, \phi$. We note that when the parameter $\lambda=\lambda_0$, the curve $\gamma$ enters from region A to region B. Thus when $R(\lambda=\lambda_0)=R_0$, $R_0$ is the area radius of the outermost shell $r_0$. The element of the induced metric on $\Sigma$ in the exterior region $(A)$ is
\begin{equation}\label{eqn12}
    ds^2_{\Sigma} = [-(|\Lambda|R^2-M)\dot v^2 + 2\dot v\dot R]d{\lambda^2} + R^2d\phi^2,
\end{equation}
where $\dot{v}=dv/d\lambda$ and $\dot{R}=dR/d\lambda$ in the above equation. Similarly, following the same procedure for interior region $(B)$, the induced metric element is
\begin{equation}\label{eqn13}
    ds^2_{\Sigma}=[-\dot{t}^2+\dot{r}^2 cos^2(\sqrt{|\Lambda|} t)]d\lambda^2 +r^2\cos^2(\sqrt{|\Lambda| }t)d\phi^2.
 \end{equation}
Here $\dot{t}$ and $\dot{r}$ are derivatives with respect to the parameter $\lambda$. Now, the maximal surface is the union of both the regions. So the volume is expressed as
\begin{equation}
\begin{split}
V_{\Sigma}[\gamma]  =- \int_{\lambda_h}^{\lambda_0} d\lambda \int_{S^1} d\phi \sqrt{R^2\big[-(|\Lambda| R^2 - M)\dot{v}^2 + 2\dot{v}\dot{R}\big]}\\- \int_{\lambda_0}^{\lambda_f} d\lambda \int_{S^1} d\phi \sqrt{r^2 cos^2(\sqrt{|\Lambda|} t)\big[-\dot{t}^2+\dot{r}^2 cos^2(\sqrt{|\Lambda|} t)\big]}.
\end{split}
\end{equation}
The minus sign is because the parameter $\lambda$ is chosen to be monotonically decreasing with increasing radius. In the first term,  $R(\lambda = 0) = R_h =\sqrt{M/|\Lambda|}$, which is the event horizon. The above expression is obtained by connecting the hypersurface across the boundary $r=r_0$. The first term in the above equation is the volume inside the BTZ black hole from the event horizon to an area radius $R(\lambda= \lambda_0) = R_0 = r_0 cos(\sqrt{|\Lambda|}t)$, which is the outer shell of the collapsing dust cloud, inside the BTZ black hole. The solution to this term where we seek the maximal volume in the interior of the vacuum part of a nonrotating BTZ black hole is solved completely in \cite{MZ}.\par
We now focus on the evolution of the maximal surface in the region (B) as a function of comoving time $t$. We follow the same Lagrangian procedure outlined in \cite{CR} for the metric given by Eq.(\ref{eqn13}). We note here that as far as the comoving coordinate chart is concerned, the location of the apparent horizon or the event horizon does not appear as any coordinate singularities in the metric. So the Euler-Lagrange equations remain the same irrespective of the boundary limits we impose. This is one of the main advantages of using the comoving coordinates to describe dust evolution. We have to maximize the following functional after integrating over the angle variable 
\begin{equation}\label{eqn15}
V_{\Sigma}  =  -2\pi \int_{\lambda_0}^{\lambda_f} d\lambda  \sqrt{r^2 cos^2(\sqrt{|\Lambda|} t)\big[-\dot{t}^2+ \dot{r}^2 cos^2(\sqrt{|\Lambda|} t)\big]}.
\end{equation}
We may choose the parameter to be $r$ since this coordinate is monotonic in the domain under consideration. The above functional then has only one function $t(r)$ that needs to be determined. The volume expression now becomes
\begin{equation}\label{eqn16}
\begin{split}
V_{\Sigma} & = - 2\pi \int_{r_0}^{0} dr  \sqrt{r^2 cos^2(\sqrt{|\Lambda|} t)\big[-\dot{t}^2+cos^2(\sqrt{|\Lambda|} t)\big]},
\end{split}
\end{equation}
where, $\dot{r}=1$ and $\dot t  =\partial t/\partial r$, is the derivative of $t$ w.r.to $r$. This can be viewed as an extremization problem and our goal is to find the equations of motion for the Lagrangian defined as 
\begin{equation}\label{eqn17}
    L = L(t, \dot{t},r) = \sqrt{r^2 cos^2(\sqrt{|\Lambda|} t)\big[-\dot{t}^2+cos^2(\sqrt{|\Lambda|} t)\big]}.
\end{equation}
The Euler-Lagrange equation is defined as 
\begin{equation}\label{eqn18}
    \frac{d}{dr}\bigg(\frac{\partial L}{\partial \dot{t}}\bigg) - \frac{\partial L}{\partial t} = 0;
\end{equation}
for the Lagrangian, (\ref{eqn17}), we  get
\begin{multline}\label{eqn19}
    \ddot{t} - \bigg(\frac{sec^2{(\sqrt{|\Lambda|}t)}}{r}\bigg)\dot{t}^3  + 3\sqrt{|\Lambda|}\big[tan(\sqrt{|\Lambda|}t)\big]\dot t^2 + \frac{\dot{t}}{r}\\ - \sqrt{|\Lambda|}\sin{(2\sqrt{|\Lambda|}t)} = 0,
\end{multline}
where $\dot t = \partial t/\partial r$ and $a'(t) = \partial a(t)/\partial t$. The solution of Eq.(\ref{eqn19}) yields the comoving time as a function of $r$, i.e., $t(r)$, which corresponds to maximal hypersurface. 

\subsection{Lagrangian formulation for the maximal volume of a black hole in D-dimensions}
In this subsection, we consider a spherically symmetric $D(=n+2)$-dimensional evolving dust cloud in the presence of a cosmological constant. The construction of the spacelike hypersurface between the event horizon and the singularity proceeds the same way as the $(2+1)$ dimensional case. We again have regions $(A)$ and $(B)$ as shown in the Fig. \ref{fig:1}, with $(A)$ being the exterior of the dust cloud within the event horizon and $(B)$ the interior of the dust cloud. We now focus on region $(B)$.  From Eqs. (\ref{metric}) and (\ref{elambda}) with $f(r)=0$, the metric element in the interior of the black hole is defined as
\begin{equation}\label{eqn32}
      ds^2 = -dt^2 + R'^2dr^2 + R^2 (t,r)~d{\Omega_n}^2.
\end{equation}
 The total volume of the black hole is the sum of the interior of dust and the exterior lying within the event horizon. Applying the conditions (\ref{eqn10}) and (\ref{eqn11}) in D(= n+2)-dimensional case we get the induced metric element as
\begin{equation}\label{eqn33}
    ds^2_{\Sigma} = ( -\dot t^2 + \dot r^2R'^2) d\lambda^2 + R^2(t,r) d{\Omega}^2_{n}.
\end{equation}
The volume in D$(= n+2)$-dimensions is defined as
\begin{equation}\label{eqn34}
\begin{split}
    V^{(D)}_{\Sigma} = \int d\lambda d\Omega_{n} \sqrt{R^{2n}(-\dot t^2 + \dot r^2R'^2)}\\ = \frac{2\pi^{\frac{n+1}{2}}}{\Gamma (\frac{n+1}{2})}\int d\lambda \sqrt{R^{2n}(-\dot t^2 + \dot r^2R'^2)}.
    \end{split}
\end{equation}
Hence, the Lagrangian is defined as
\begin{equation}\label{eqn35}
    L(t,\dot t, \lambda) = \sqrt{R^{2n}(-\dot t^2 + \dot r^2R'^2)}.
\end{equation}
Now, let the parameter $\lambda = r$ then $\dot{r}= 1$, substitute $R{(t,r)} = r a(t)$ and $R'(t,r) = a(t)$ in Eq.(\ref{eqn35}),where $a(t)$ is the scale parameter, and we get
\begin{equation}\label{eqn36}
    L(t,\dot{t}, r) = \sqrt{[ra(t)]^{2n}(-\dot{t}^2 + [a(t)]^2)}.
\end{equation}
Our next goal is to find the equation of motion for the above Lagrangian from Eq.(\ref{eqn18}); we get
\begin{multline}\label{eqn37}
   \ddot{t} - \bigg(\frac{n}{r[a(t)]^2}\bigg){\dot{t}}^3 - \bigg(\frac{(n+2)a'(t)}{a(t)}\bigg)\dot t^2 + \bigg(\frac{n}{r}\bigg)\dot{t}\\
   + (n+1)a(t)a'(t) = 0,
\end{multline}
where $\dot t = \partial t/\partial r$ and $a'(t) = \partial a(t)/\partial t$. This is the differential equation in D-dimensions, and we can easily recover the $(2+1)$-dimensional case by substituting $n = 1$ and $a(t) = \cos(\sqrt{|\Lambda|}t)$ in Eq.(\ref{eqn37}). For estimating the maximal hypersurface in the trapped region, we can choose the appropriate boundary values. We note that the condition for marginal trapping for the model under consideration is given by $r_a=-1/\dot{a}$ \cite{KS}, where the $\dot{a}$ is a derivative of the scale factor with respect to time $t$. In a collapsing scenario, we have $\dot{a}$ is negative thereby giving a positive value for $r_a$.
 
\section{ Reinhart radius in (2 + 1) dimensional evolving dust model}
 In this section, we discuss the existence of Reinhart radii in various evolving dust models in $(2+1)$-dimensions. As discussed in the Introduction, these are special radii that have been found in the interior region of black holes. In the paper \cite{Reinhart}, Reinhart showed that in the interior of the Schwarzschild black hole, the hypersurface $R_R=3M/2$ is a maximal hypersurface. He used the vanishing of the trace of the extrinsic curvature to arrive at this value for the radius. As pointed out in the Introduction, this radius has proved useful in approximating the interior volume of a black hole in the paper \cite{CR}. This radius was found in \cite{CR} via an independent method whereby the volume inside the Schwarzschild black hole was maximized. Similarly in the papers  \cite{XY,SSR,BJ,BZ,MZ,CL,Bibhas} the maximization of the internal volume yielded the Reinhart radius. All these models describe static or stationary black holes. The Reinhart radii in the more general setting is a gap in the literature. This is the goal of the second part of the article. We now track the existence of the Reinhart radii in various cases in $(2+1)$-dimensions. The location of the Reinhart radius and its correlation with the evolution of the apparent horizon can provide important clues toward estimating the volume of the evolving black hole. To this end, we evaluate the divergence of normal to $R(t,r)=const.$ surface. To find the normal to the surface we differentiate $R(t,r)=const.$ and get 
\begin{equation}\label{constR17}
 \dot{R}dt + R'dr = 0.
\end{equation}
The covariant components of normal vector defined as $n_{\alpha}=(n_{t},n_{r})$, where $n_{t}= \dot{R}$ and $n_{r}= R'$. We now analyze the case of $(2+1)$-dimensional scenario separately here owing to the nontrivial nature of gravity in $(2+1)$-dimensions. We start with the area radius $R(t,r)$ of the comoving shell as a function of the comoving time and shell label $r$. The normalized contravariant components of normal vector are $N^{\alpha}=(N^{t},N^{r}) = \bigg(\frac{-\dot{R}}{\sqrt{\dot{R}^2-1}} , \frac{1}{R'\sqrt{\dot{R}^2-1}}\bigg)$. The condition for the vanishing trace of extrinsic curvature, which implies that the normal vector is divergence free,  is written as
\begin{equation}\label{nalpha18}
    N^{\alpha}_{;\alpha} = \frac{1}{\sqrt{-g^{(3)}}} {\frac{\partial}{\partial x^{\alpha}}({\sqrt{-g^{(3)}} N^{\alpha})}} = 0,
\end{equation}
where $\sqrt{-g^{(3)}} = R'R$, which is obtained from the determinant of metric (\ref{eqn32}) for $n=1$ i.e, $(2+1)$-dimensional case. Equation (\ref{nalpha18}) can be written as
\begin{equation}
    \frac{1}{R'R} \bigg[{\frac{\partial}{\partial t}({R'R N^{t})}} + {\frac{\partial}{\partial r}({R'R N^{r})}}\bigg] = 0
\end{equation}
or
\begin{equation}\label{partialnalpha20}
    -\frac{\partial}{\partial t}\bigg(\frac{\dot{R} R' R}{\sqrt{\dot{R}^2-1}}\bigg) + {\frac{\partial}{\partial r}\bigg(\frac{R}{\sqrt{\dot{R}^2-1}}\bigg)} = 0,
\end{equation}
After simplifying Eq.(\ref{partialnalpha20}), we get
\begin{equation}\label{rprime21}
    (\dot{R}^2-1)^2 + \frac{R\dot{R}^3\dot{R}'}{R'} - R \ddot{R} = 0,
\end{equation}    
where the expression for $\dot{R}^2$ is obtained for $(2+1)$-dimensions by substituting $n = 1$ into Eq.(\ref{rdotsquare}), and we get
\begin{equation}\label{rdotsqure}
    \dot{R}^2 = f(r) + \Lambda R^2 + F(r).
\end{equation}
We take the $f(r) = 0$ for the marginally bounded shell of the dust cloud. Now differentiating Eq.(\ref{rdotsqure}) with respect to $t$ and $r$, we get expression for $\ddot{R}$ and $\dot{R}'$ as
\begin{equation}\label{rdoubledot}
    \ddot{R} = \Lambda R
\end{equation}
and
\begin{equation}\label{rdotprime}
    \dot{R}' = \frac{\Lambda R R'}{\dot{R}} + \frac{F'(r)}{2\dot{R}}. 
\end{equation}
Substituting the value of $\dot{R^2}, \ddot{R}$, and $\dot{R}'$ from Eqs.(\ref{rdotsqure}), (\ref{rdoubledot}) and (\ref{rdotprime}) into Eq.(\ref{rprime21}), we get
\begin{equation}\label{fprime25}
    \frac{RF'(r)}{2R'}[\Lambda R^2+F(r)]+[{\Lambda}R^2 + F(r) -1][2\Lambda R^2 + F(r) - 1]  = 0.
\end{equation}
 We now obtain a general formula connecting the comoving energy density at a point with the area radius $R(t,r)$, Misner-Sharp mass $F(r)$, and cosmological constant $\Lambda$. Let $\epsilon$ be the energy density of the collapsing dust cloud then we can define the energy density as $\epsilon =\frac{ F'(r)}{2 R R'}$. Therefore, from Eq.(\ref{fprime25}), we can write $\epsilon$ as
\begin{equation}
    \epsilon = \frac{ F'(r)}{2 R R'} = -\frac{[{\Lambda}R^2 + F(r) -1][2\Lambda R^2 + F(r) - 1]}{R^2 [\Lambda R^2 + F(r)]}.
\end{equation}
This is the condition for $R=const.$ hypersurface to have a vanishing trace of extrinsic curvature. Though this formula is true in general, we now examine a few simple cases and gain an understanding of this special point in the hypersurface.
\subsection{Vacuum case} 
When $F(r)=0$. This situation represents matter content is zero in the entire spacetime. We see that 
\begin{equation}\label{eqroot33}
   [{\Lambda}R^2  -1][2\Lambda R^2  - 1] = 0.
\end{equation}
We have two values of area radius where the condition is met: $R_{CH}=1/\sqrt{\Lambda}$ and $R_{R}=1/\sqrt{2\Lambda}=R_{CH}/\sqrt{2}.$ The former root is the de Sitter cosmological horizon, which is a null horizon. The normal to the horizon is also the null generator and has vanishing divergence. We look for timelike normals  (timelike normals for $R=const$ occur in the interior of a black hole)  and spacelike normals, so this root is not our answer. The other root is the answer and we observe that this root lies between the $R=0$ and the cosmological horizon $R_{CH}$. This feature is also observed in the situations to come.  We note that for a negative cosmological constant, this case does not yield any solutions.
\subsection{Static black hole case }
Suppose we consider the case when the Misner-Sharp mass $F(r)$ is a constant, this is the situation of a black hole or a naked singularity where all the mass has already collapsed to a point. We can use the formula derived above to examine the situation of a black hole using the comoving chart and not static/stationary coordinates. So we take the condition that $F(r)=const.=M+1$ where $M$ is the ADM mass. Also, we note that automatically $F'(r)=0$. From Eq.(\ref{fprime25}), we get
\begin{equation}\label{eqroot34}
   [{\Lambda}R^2 + M ][2\Lambda R^2 + M] = 0;
\end{equation}
there are again two roots of the Eq.(\ref{eqroot34}),  
\begin{equation}
    R =R_{CH} = \sqrt{\frac{-M}{\Lambda}}\ and \ R = R_{R} =\sqrt{\frac{-M}{2\Lambda}}.
\end{equation}
Here we have two subcases. For a positive cosmological constant, we require that the mass function $F(r)$ at the naked singularity (which is a conical singularity) is not greater than 1 so that $M$ is negative. In this case, we have the cosmological horizon at $R_{CH}$, and as can be seen in the above equation, $R_{R}=R_{CH}/\sqrt{2}$. The fact that $M$ has to be negative is observed in \cite{rossandmann,sashideep}. This scenario is not present in the four and higher dimensions where the ADM mass $M$ is positive definite. Here $R_{R} < R_{CH}$, which shows that the surface is not beyond the cosmological horizon. The second subcase involves a negative cosmological constant. Here, in contrast, we require $M$ to be positive for roots to be real. So $F>1$ for a black hole event horizon to exist. The total mass that collapses has to be greater than 1 so that an event horizon forms \cite{rossandmann,sashideep}.  We then have 
 \begin{equation}
   R =R_{e} = \sqrt{\frac{M}{|\Lambda|}}\ and \ R = R_{R} =\sqrt{\frac{M}{2|\Lambda|}}.
\end{equation}
As is evident from the above equations, $R_{R}$ occurs in the interior of the BTZ black hole with zero angular momentum. The radius is $R_{R}=R_{e}/\sqrt{2}$. This is similar to the situation of the case of four-dimensional black holes as observed in the works of \cite{CR,BZ,XY,CL}.
\subsection{Cosmological case}
We look at cosmological solutions in $(2+1)$-dimensions with and without a cosmological constant. The mass function for homogeneous collapsing dust in $(2+1)$-dimensions are defined as $F(r) = g r^2$, and the area radius of comoving shells is $R = R(t,r) = r a(t)$. Substituting these parameters in Eq.(\ref{fprime25}), we get
\begin{equation}\label{eqroot29}
    2(\Lambda a^2 +g )^2r^4 - (3\Lambda a^2 + 2g) r^2 + 1 = 0.
\end{equation}
In case (1) when $\Lambda = 0$, i.e., without cosmological constant, we set $\Lambda=0$ in the above equation then we obtain, $g^2r^4+(gr^2-1)^2=0$. As can be seen readily, this is a sum of squares and is never zero. So we do not have $R_{R}$ for real values of the area radius. To analyze the case with a nonzero cosmological constant, we evaluate the roots of Eq.(\ref{eqroot29}). This gives Reinhart radius $r_{R}$  as
\begin{equation}\label{rss30}
    r_{R} = \bigg(\frac{(3\Lambda a^2 + 2g) \pm \sqrt{(3\Lambda a^2 + 2g)^2-8(\Lambda a^2 +g)^2}}{4(\Lambda a^2 + g )^2}\bigg)^{\frac{1}{2}}.
\end{equation}
In case (2) when $\Lambda >0$, i.e.,  de Sitter spacetime $r_{R}$ will be positive only when the terms inside the square root are positive, i.e,
\begin{equation} \label{lambda31}
\begin{split}
    (3\Lambda a^2 + 2g)^2-8(\Lambda a^2 +g)^2 >0  \\ 
    \Rightarrow a^2 > -\frac{2g(\sqrt{2}+1)}{{\Lambda(3+2\sqrt{2})}} \ and \ a^2 > \frac{2g(\sqrt{2}-1)}{{\Lambda(3-2\sqrt{2})}} \\
    \Rightarrow a^2 > \frac{2g(\sqrt{2}-1)}{{\Lambda(3-2\sqrt{2})}} \ or \  a^2>\frac{2g(\sqrt{2}+1)}{\Lambda}.
        \end{split}
\end{equation}
This condition can be seen to be equal to the relation inside the square root has to be positive. This inequality implies that $r_{R}$ is only possible during the evolution if, the scale parameter is greater than a critical value given in the equation. This is a surprising fact that this point emerges during the course of evolution provided the scale parameter crosses a certain threshold value. \\
In case (3) when $\Lambda <0$, i.e., anti-de Sitter spacetime, we replace $\Lambda$ with $-\Lambda$ in Eq.(\ref{rss30}), and we get
\begin{equation}\label{rss30'}
   r_{R} = \bigg(\frac{(-3|\Lambda| a^2 + 2g) \pm \sqrt{\Lambda^2 a^4 + 4|\Lambda| g a^2 - 4g^2)}}{4(-|\Lambda| a^2 + g )^2}\bigg)^{\frac{1}{2}}
\end{equation}
Let us define $A = -3|\Lambda| a^2 + 2g$ and $B= \sqrt{8}(-|\Lambda| a^2 +2g)$, then the terms, $A^2-B^2= (-3|\Lambda| a^2 + 2g)^2 - 8(\Lambda a^2 +g)^2 = \Lambda^2 a^4 + 4|\Lambda| g a^2 - 4g^2$. There are two possible situations that arise on $A$:\\
(i). Suppose $A$ is negative, i.e, $A<0$, then Eq.(\ref{rss30'}) can be written as
\begin{equation}\label{eqn41'}
    r_{R} = \bigg(\frac{-|(-3|\Lambda| a^2 + 2g)| \pm \sqrt{\Lambda^2 a^4 + 4|\Lambda| g a^2 - 4g^2}}{4(-|\Lambda| a^2 + g )^2}\bigg)^{\frac{1}{2}}.
\end{equation}
Now we check the existence of the above Reinhart radius $r_{R}$ as follows:
\begin{equation}\label{eqn42'}
    \begin{split}
        (-3|\Lambda| a^2 + 2g)^2 > (-3|\Lambda| a^2 + 2g)^2 - 8(\Lambda a^2 +g)^2\\
        \Rightarrow  (-3|\Lambda| a^2 + 2g)^2 >\Lambda^2 a^4 + 4|\Lambda| g a^2 - 4g^2\\
     \Rightarrow |(-3|\Lambda| a^2 + 2g)| > \sqrt{\Lambda^2 a^4 + 4|\Lambda| g a^2 - 4g^2}\\
      \Rightarrow -|(-3|\Lambda| a^2 + 2g)| < - \sqrt{\Lambda^2 a^4 + 4|\Lambda| g a^2 - 4g^2}\\
      \Rightarrow -|(-3|\Lambda| a^2 + 2g)| + \sqrt{\Lambda^2 a^4 + 4|\Lambda| g a^2 - 4g^2} <0
    \end{split}
\end{equation}
and $ -|(-3|\Lambda| a^2 + 2g)| - \sqrt{\Lambda^2 a^4 + 4|\Lambda| g a^2 - 4g^2} <0$ is also possible. Hence, these conditions show that the terms $-|(-3|\Lambda| a^2 + 2g)| \pm \sqrt{\Lambda^2 a^4 + 4|\Lambda| g a^2 - 4g^2}<0$ in the Eq.(\ref{eqn41'}), which means $r_{R}$ becomes imaginary, and hence there is no any $r_{R}$ exist.\\
(ii). Suppose $A$ is positive, i.e., $A=(-3|\Lambda| a^2 + 2g)>0$, then it gives the condition on the scale parameter $a(t)$ as $a^2<2g/3|\Lambda|$ and using this condition inside the square root term $\sqrt{\Lambda^2 a^4 + 4|\Lambda| g a^2 - 4g^2)}$ we get an imaginary value. Hence in both scenarios, there is no solution for $r_{R}$,  and hence they do not exist for $(2+1)$-dimensional AdS cosmological spacetimes.

\section{Reinhart radii in D-dimensional evolving dust in the presence of cosmological constant}
We now extend our results in $(2+1)$-dimensions to general D-dimensions. We consider the D-dimensional metric found in Sec. II. We repeat the analysis done in a $(2+1)$-dimensional case. We find the normal to the surface $R(t,r)=const.$ and get 
\begin{equation}\label{constR17'}
 \dot{R}dt + R'dr = 0.
\end{equation}
The normalized contravariant components of normal vector can be found to be $N^{\alpha}=(N^{t}, N^{r}) = \bigg(\frac{-\dot{R}}{\sqrt{\dot{R}^2-1}}, \frac{1}{R'\sqrt{\dot{R}^2-1}}\bigg)$. The condition for the vanishing trace of extrinsic curvature in D$(= n+2)$-dimensions is given by,
\begin{equation}\label{eqn39}
    N^{\alpha}_{;\alpha} = \frac{1}{\sqrt{-g^{(D)}}} {\frac{\partial}{\partial x^{\alpha}}({\sqrt{-g^{(D)}} N^{\alpha})}} = 0,
\end{equation}
where $\sqrt{-g^{(D)}} = R'R^n {\Theta}_{n}$, which is obtained from the determinant of metric (\ref{eqn32}), where ${\Theta}_{n}$ contains the product of all the angular parts of determinant of metric (\ref{eqn32}). Equation (\ref{eqn39}) can be written as
\begin{equation}\label{eqn40}
    \frac{1}{R'R^n {\Theta}_{n}} \bigg[{\frac{\partial}{\partial t}({R'R^n {\Theta}_{n} N^{t})}} + {\frac{\partial}{\partial r}({R'R^n {\Theta}_{n} N^{r})}}\bigg] = 0
\end{equation}
or
\begin{equation}\label{eqn41}
    -\frac{\partial}{\partial t}\bigg(\frac{\dot{R} R' R^n}{\sqrt{\dot{R}^2-1}}\bigg) + {\frac{\partial}{\partial r}\bigg(\frac{R^n}{\sqrt{\dot{R}^2-1}}\bigg)} = 0.
\end{equation}
After simplifying Eq.(\ref{eqn41}), we get
\begin{equation}\label{eqn42}
    n R^{n-1}(\dot{R}^2-1)^2 + \frac{R^n\dot{R}^3\dot{R}'}{R'} - R^n \ddot{R} = 0.
\end{equation}
Now, differentiating Eq.(\ref{rdotsquare}) with respect to $t$ and $r$, we get the expression for $\ddot{R}$ and $\dot{R}'$ as
\begin{equation}\label{eqn44}
    \ddot{R} = \frac{2\Lambda R}{n(n+1)} - \frac{(n-1)F(r)}{2 R^{n}}
\end{equation}
and
\begin{equation}\label{eqn45}
    \dot{R}' = \frac{2\Lambda R R'}{n(n+1)\dot{R}} + \frac{F'}{2R^{n-1}\dot{R}} - \frac{(n-1) F R'}{2 R^n\dot{R}}.
\end{equation}
Substituting the value of $\dot{R^2}, \ddot{R}$, and $\dot{R}'$ from Eqs.(\ref{rdotsquare}), (\ref{eqn44}) and (\ref{eqn45}) into Eq.(\ref{eqn42}) and after simplifying, we get
\begin{multline}\label{eqn46}
    \frac{R F'(r)}{2R'}\bigg(\frac{2\Lambda R^2}{n(n+1)}+ \frac{F(r)}{R^{n-1}}\bigg)+\\ \bigg(\frac{2\Lambda R^2}{n(n+1)} + \frac{F(r)}{R^{n-1}}-1\bigg) \bigg[\frac{2\Lambda}{n}R^{n+1} + \frac{(n+1)}{2}F(r) - nR^{n-1} \bigg]\\ = 0.
\end{multline}
We will see below that the above formula simplifies to an expression involving coordinate invariant since the term containing $F'$ gets related to the energy density. The formula therefore is an interesting relation between the principle value of the energy-momentum tensor, the cosmological constant, and the Misner-Sharp mass.\par
In a general setting, the Misner-Sharp mass is a monotonically increasing function of the comoving radius $r$. We take a nonzero value for the mass function $F(r)\neq 0$ (and $F'(r)\neq 0$). Let $\epsilon$ be the energy density of collapsing dust then we can define the energy density as $\epsilon = \frac{n F'(r)}{2 R^n R'}$. Therefore, from Eq.(\ref{eqn46}), we can write $\epsilon$ as
\begin{multline}\label{eqn47}
    \epsilon = \frac{n F'(r)}{2 R^n R'} =\\ \frac{n\bigg(1 - \frac{2\Lambda R^2}{n(n+1)} - \frac{F(r)}{R^{n-1}}\bigg) \bigg[\frac{2\Lambda}{n}R^{n+1} + \frac{(n+1)}{2}F(r) - nR^{n-1}\bigg] }{R^{n+1} \bigg(\frac{2\Lambda R^2}{n(n+1)} + \frac{F(r)}{R^{n-1}}\bigg)}.
\end{multline}
\subsection{Vacuum scenario} 
When there is no matter in the spacetime we have $F(r)=0$. In this case, the above equation yields Minkowski, de Sitter, or anti-de Sitter spacetime. For the case of $F(r)=0$, the formula above gives
    \begin{multline}\label{Fequalszero}
    \\ \bigg(\frac{2\Lambda R^2}{n(n+1)} -1\bigg) \bigg[\frac{2\Lambda}{n}R^{n+1}  - nR^{n-1} \bigg] = 0.
\end{multline}
 We solve for the $R_{R}$,
\begin{equation}\label{eqn56}
     \frac{2\Lambda R^2}{n(n+1)} -1 = 0
\end{equation}
and
\begin{equation}\label{eqn57}
     \frac{2\Lambda}{n} R^{n+1} - nR^{n-1}=0
\end{equation}
from Eqs.(\ref{eqn56}) and (\ref{eqn57}), we get
\begin{equation}\label{eqn58}
    R_{CH} = \sqrt{\frac{n(n+1)}{2\Lambda}}\ and \ R_{R} = \sqrt{\frac{n^2}{2\Lambda}}
\end{equation}    
or
\begin{equation}\label{eqn59}
    R_{R} = \sqrt{\frac{n}{n+1}}R_{CH}.
\end{equation}
 Here $n<n+1$ so $R_{R}<R_{CH}$, which means Reinhart radius occurs before the cosmological horizon. The cosmological horizon is the boundary of the antitrapped region. So $R_{R}$ occurs in the accessible part of spacetime from the point of view of the interior of the cosmological horizon.  We also note that this surface is possible only in de Sitter (dS) space and not possible in anti-de Sitter (AdS) space.
\subsection{D-dimensional Schwarzschild black hole } 
When $\Lambda = 0$ and mass function $F(r) = const.$ then $F'(r) = 0$. From Eq.(\ref{eqn46}), we get
\begin{equation}\label{eqn48}
    \bigg(\frac{F(r)}{R^{n-1}}-1\bigg)\bigg[\frac{n+1}{2}F(r) - nR^{n-1}\bigg] = 0
\end{equation}
or
\begin{equation}\label{eqn49}
    R = R_e =[F(r)]^{\frac{1}{n-1}}\ and \  R = R_{R} = \bigg(\frac{n+1}{2n}F(r)\bigg)^{\frac{1}{n-1}}.
\end{equation}
From Eq.(\ref{eqn49}), we can write
\begin{equation}\label{eqn51}
    R_{R} = \bigg(\frac{n+1}{2n}\bigg)^{\frac{1}{n-1}}R_e .
\end{equation}
For Schwarzschild case $n = 2$ (i.e., 4 dimensions), the mass function $F(r) = 2M$, then from equation (\ref{eqn49}) we get $R_e = 2M$, which corresponds to the event horizon and $R_{R} = \frac{3}{2}M$, which is the Reinhart radius that corresponds to maximal hypersurface as identified in  \cite{CR}. From Eq.(\ref{eqn51}) $R_{R} = \frac{3}{4}R_e$, i.e., $R_{R}< R_e$, which means $R_{R}$ lies inside the event horizon of a black hole since $2n>n+1$ for $n = 2,3...$. This means $R_{R}$ always lies inside the event horizon for all $n>1$.  For the case $n=1$ there is no black hole if there is no negative cosmological constant.
\subsection{D-dimensional Schwarzschild-de Sitter/anti-de Sitter spacetime scenario}
\begin{figure*}
  \subfigure[]{ \includegraphics[width=0.31\textwidth,clip]{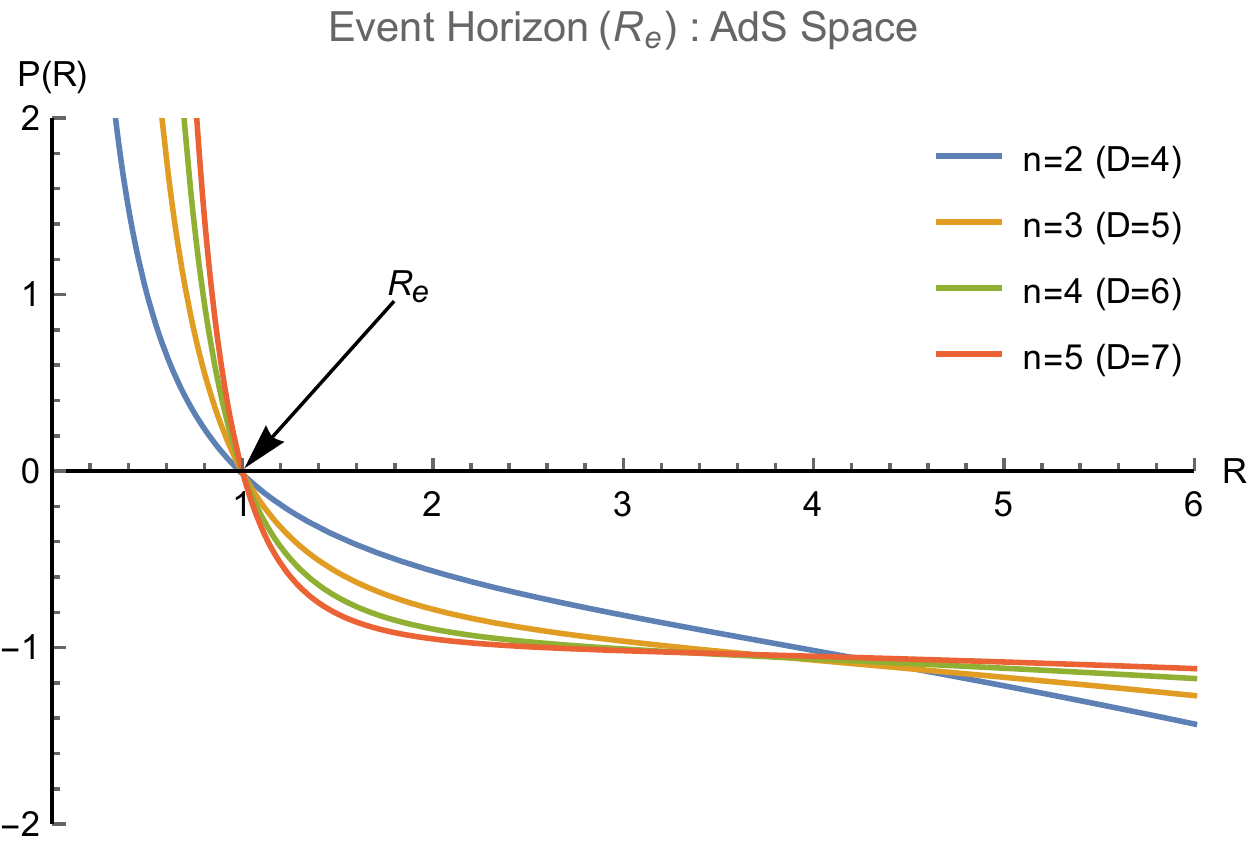} }
  \subfigure[]{ \includegraphics[width=0.31\textwidth,clip]{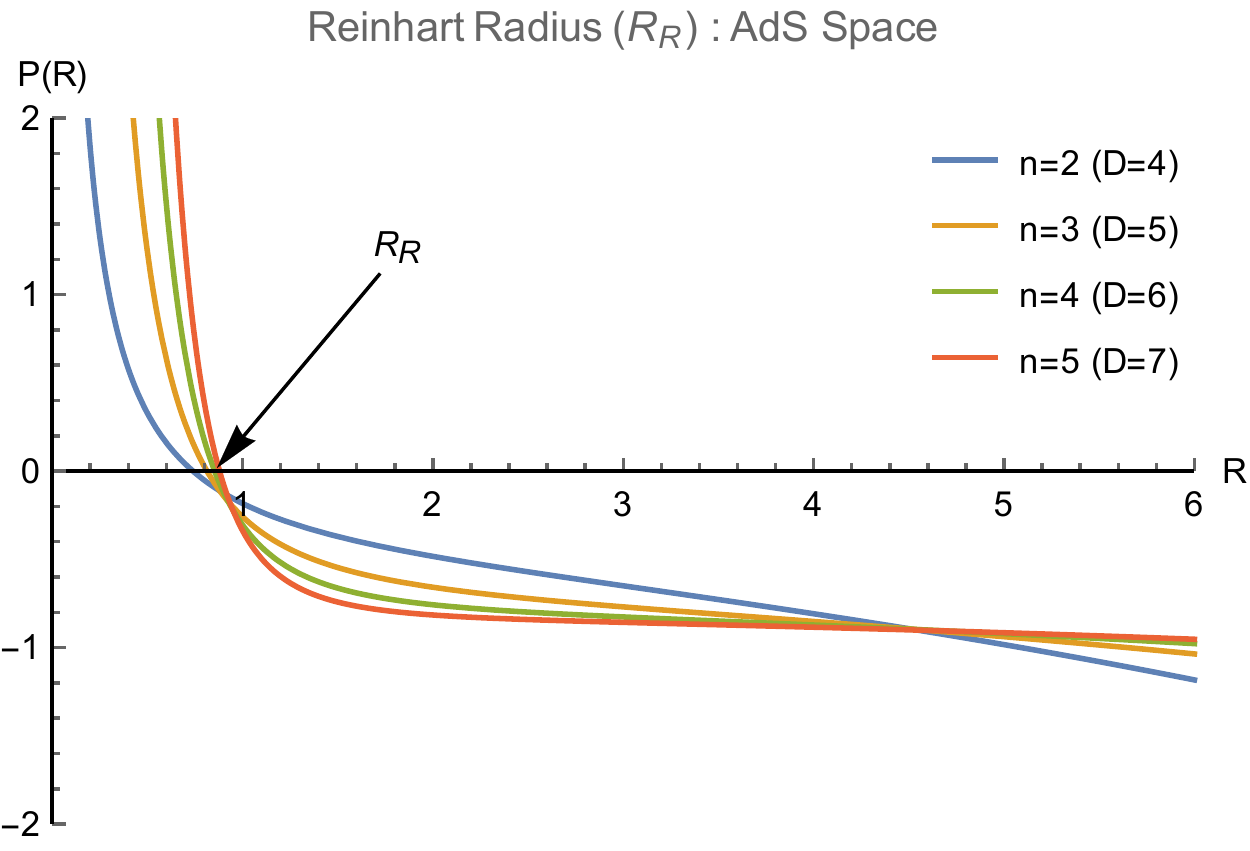} }
  \subfigure[]{ \includegraphics[width=0.31\textwidth,clip]{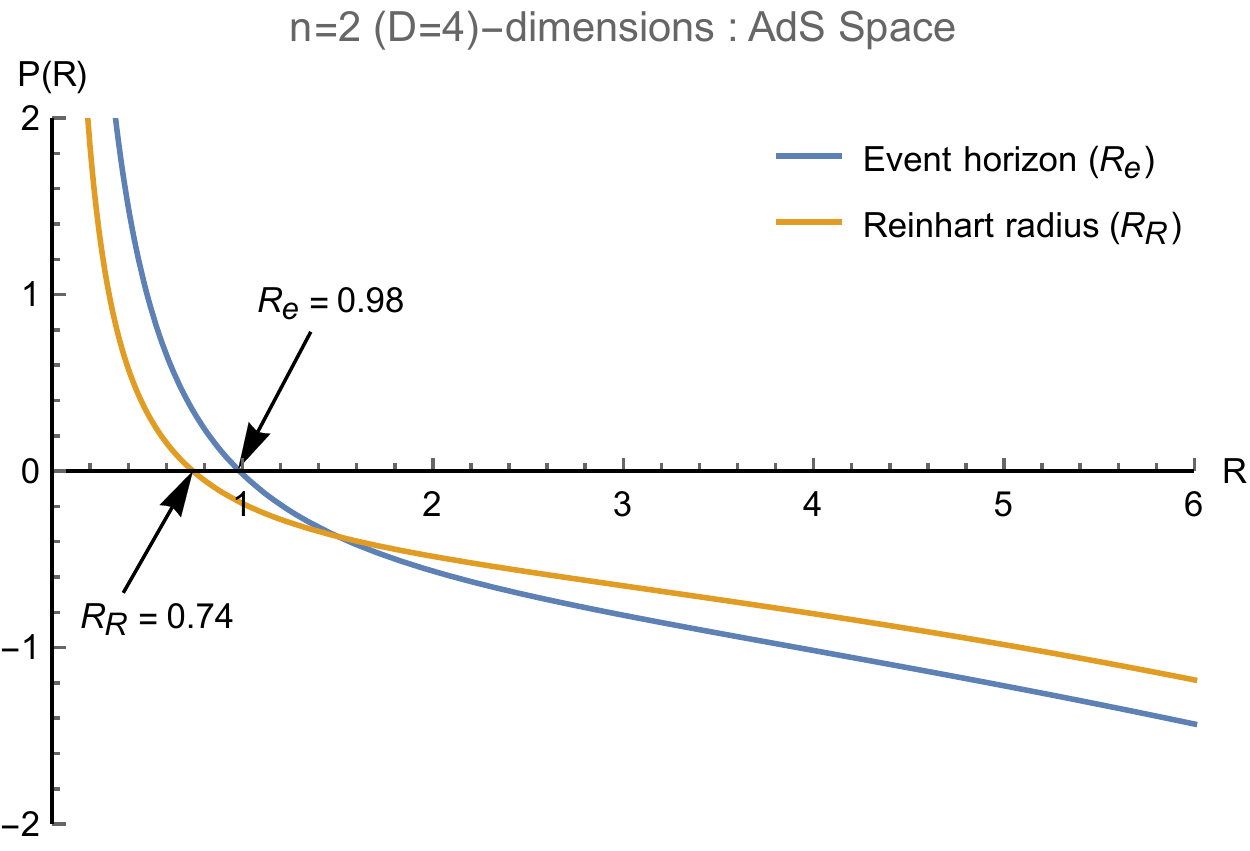} }
  \subfigure[]{ \includegraphics[width=0.31\textwidth,clip]{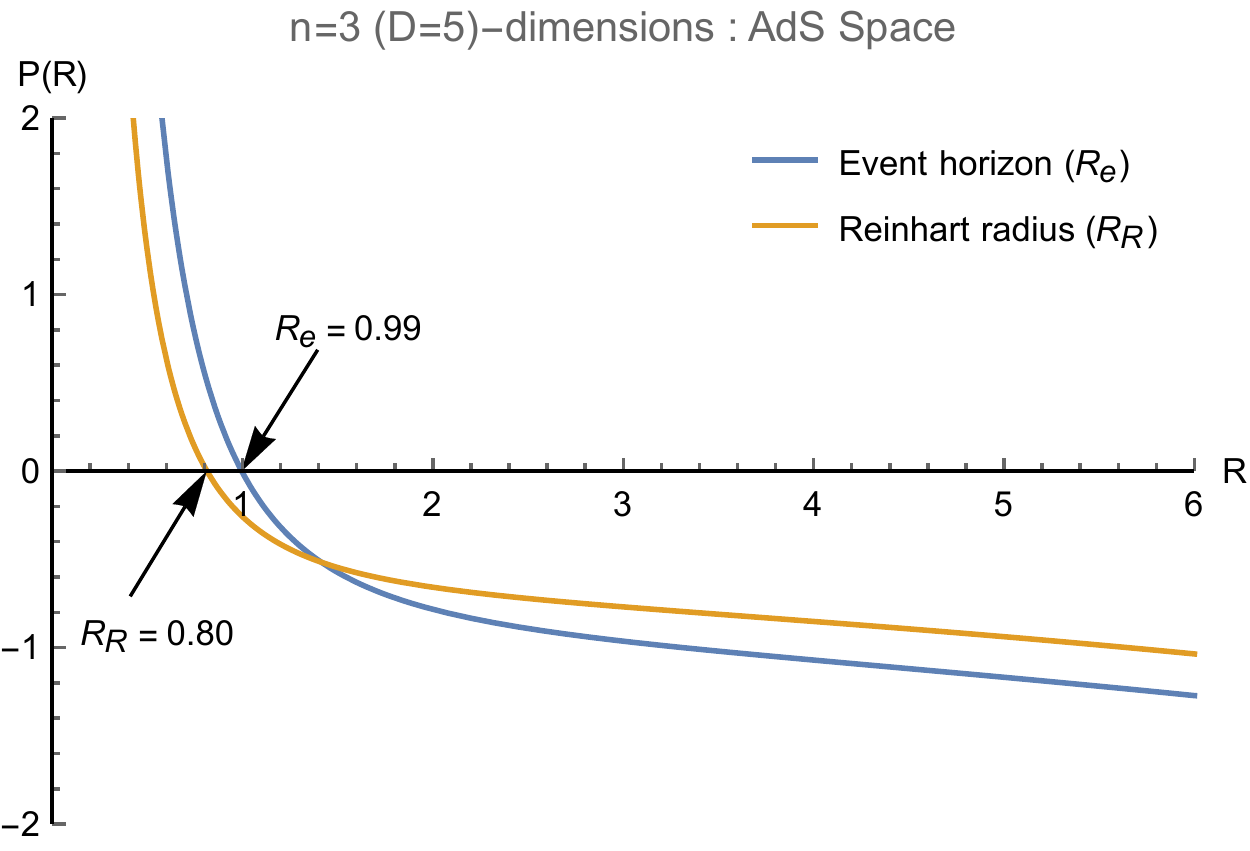} }
  \subfigure[]{ \includegraphics[width=0.31\textwidth,clip]{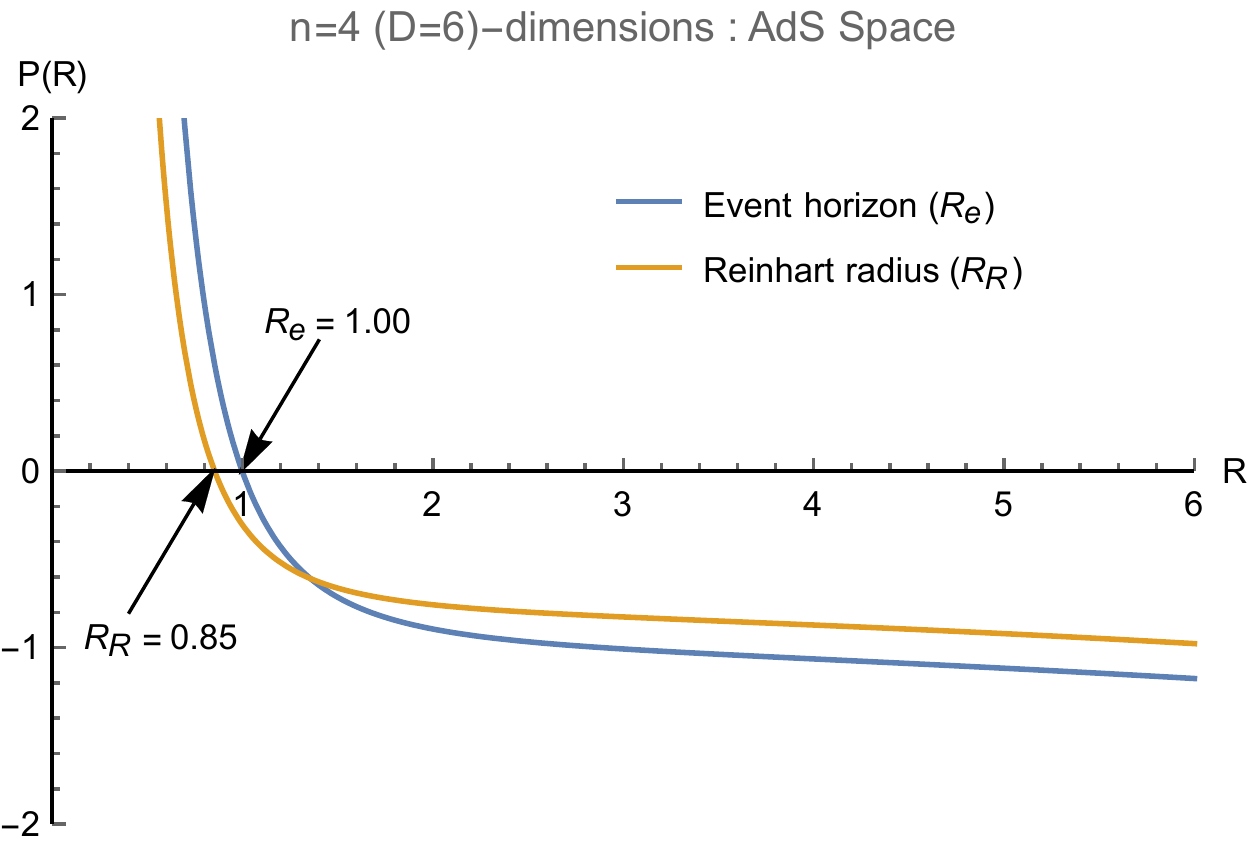} }
  \subfigure[]{ \includegraphics[width=0.31\textwidth,clip]{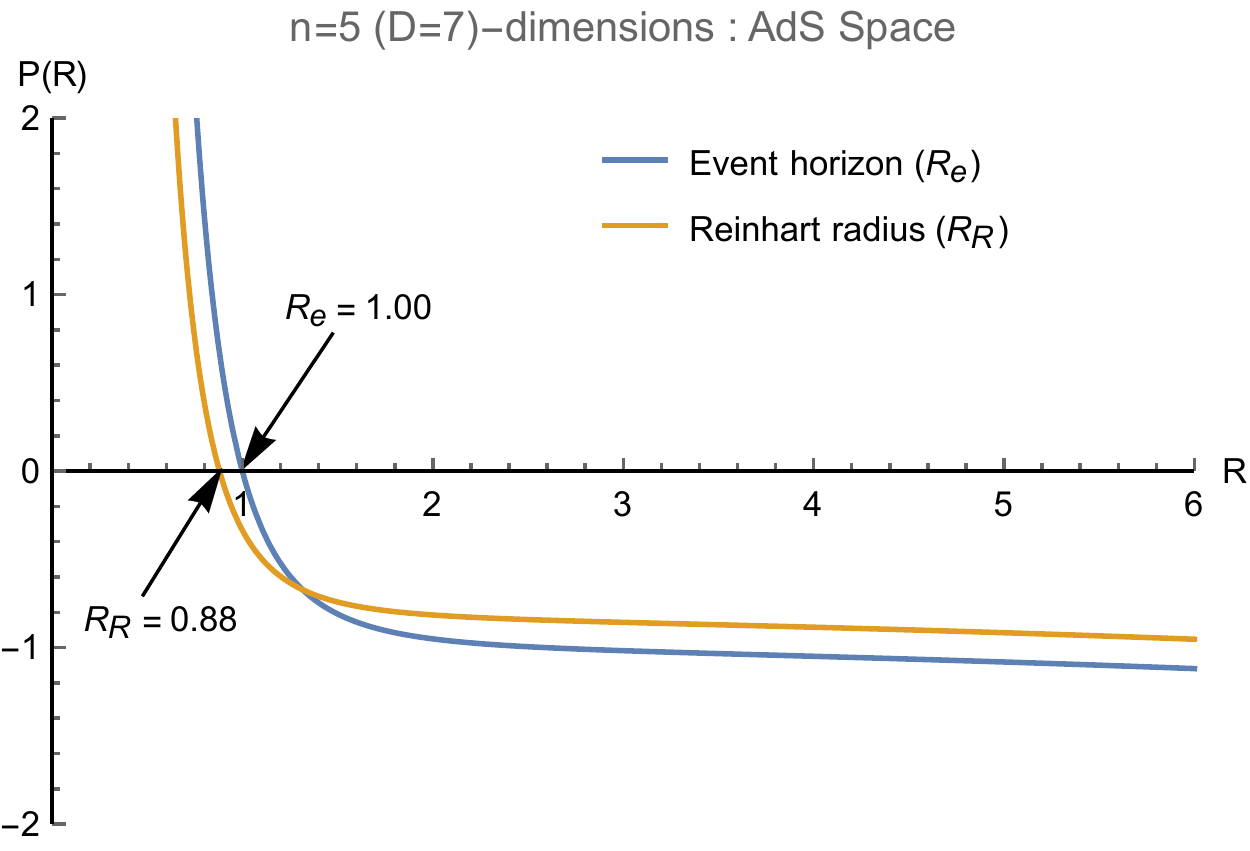} }
\caption{\label{fig:2} {The graph (a) shows the location of all the event horizons corresponding to the polynomial $P(R = R_e)$ of Eq.(\ref{eqn53}), and graph (b) shows the location of all the Reinhart radii corresponding to the polynomial $P(R=R_{R})$ of Eq.(\ref{eqn54}) for D = 4, 5, 6, and 7 dimensions in AdS spacetime. Graphs (d),(e), and (f) show that the Reinhart radius $R_{R}$ always lies inside the event horizon $R_e$. Here we take the cosmological constant $\Lambda = -0.05$ and mass function $F(r) = 1$ for all the graphs}.}
\end{figure*}

\begin{figure*}
  \subfigure[]{ \includegraphics[width=0.31\textwidth,clip]{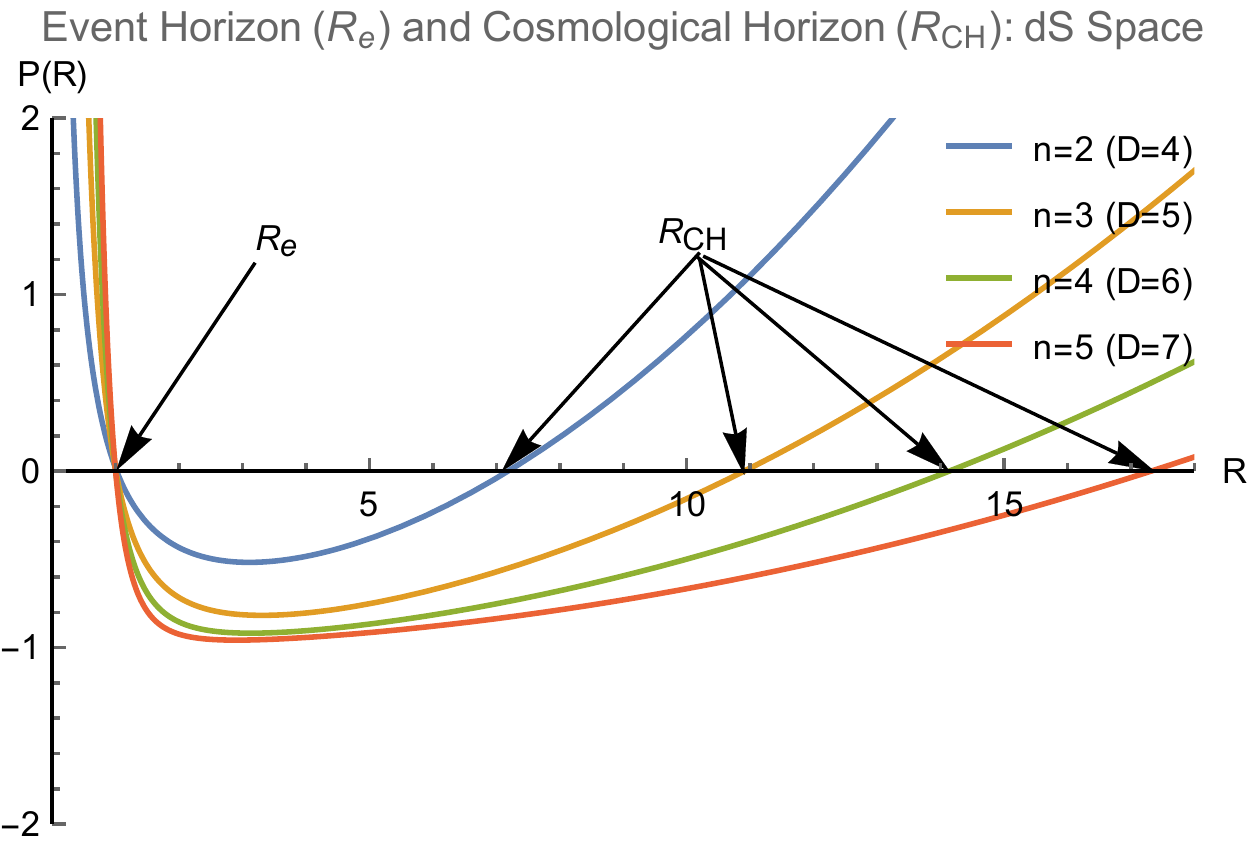} }
  \subfigure[]{ \includegraphics[width=0.31\textwidth,clip]{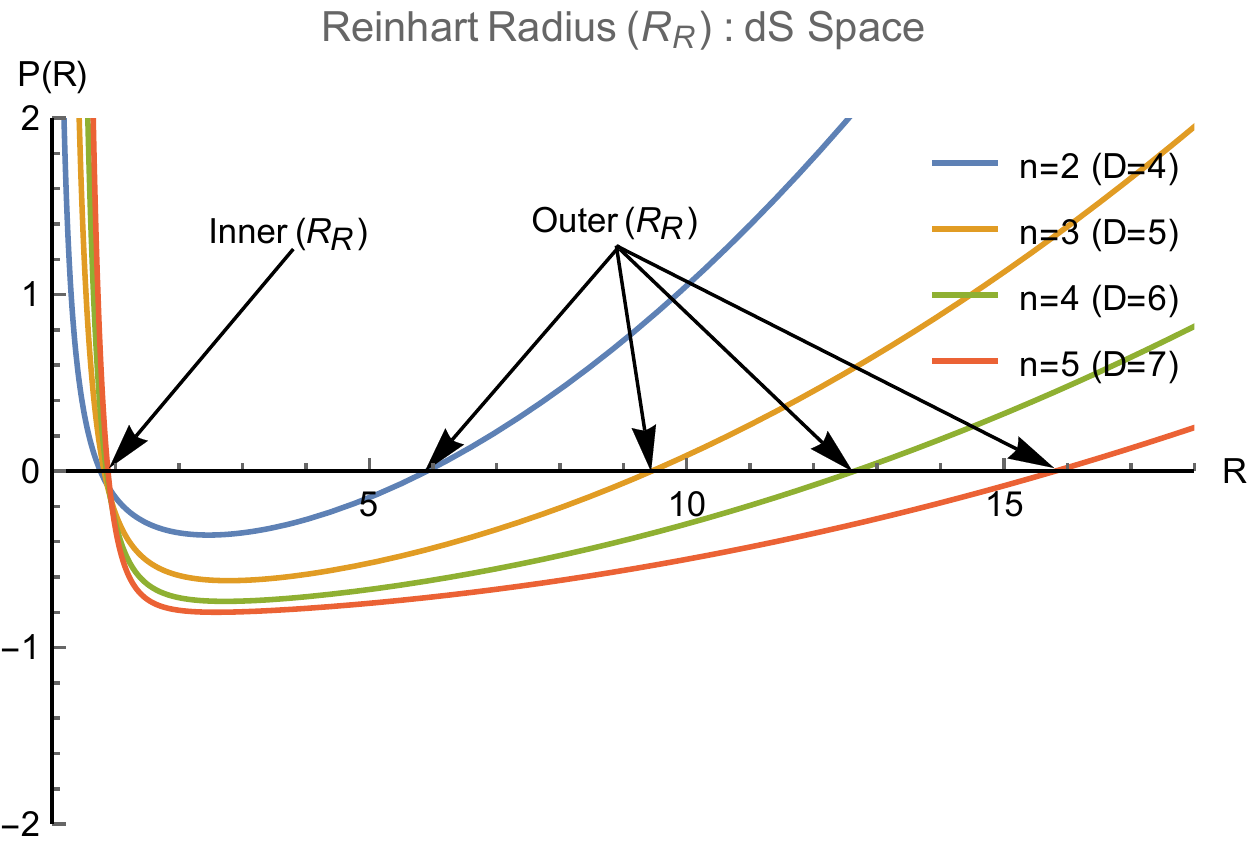} }
  \subfigure[]{ \includegraphics[width=0.31\textwidth,clip]{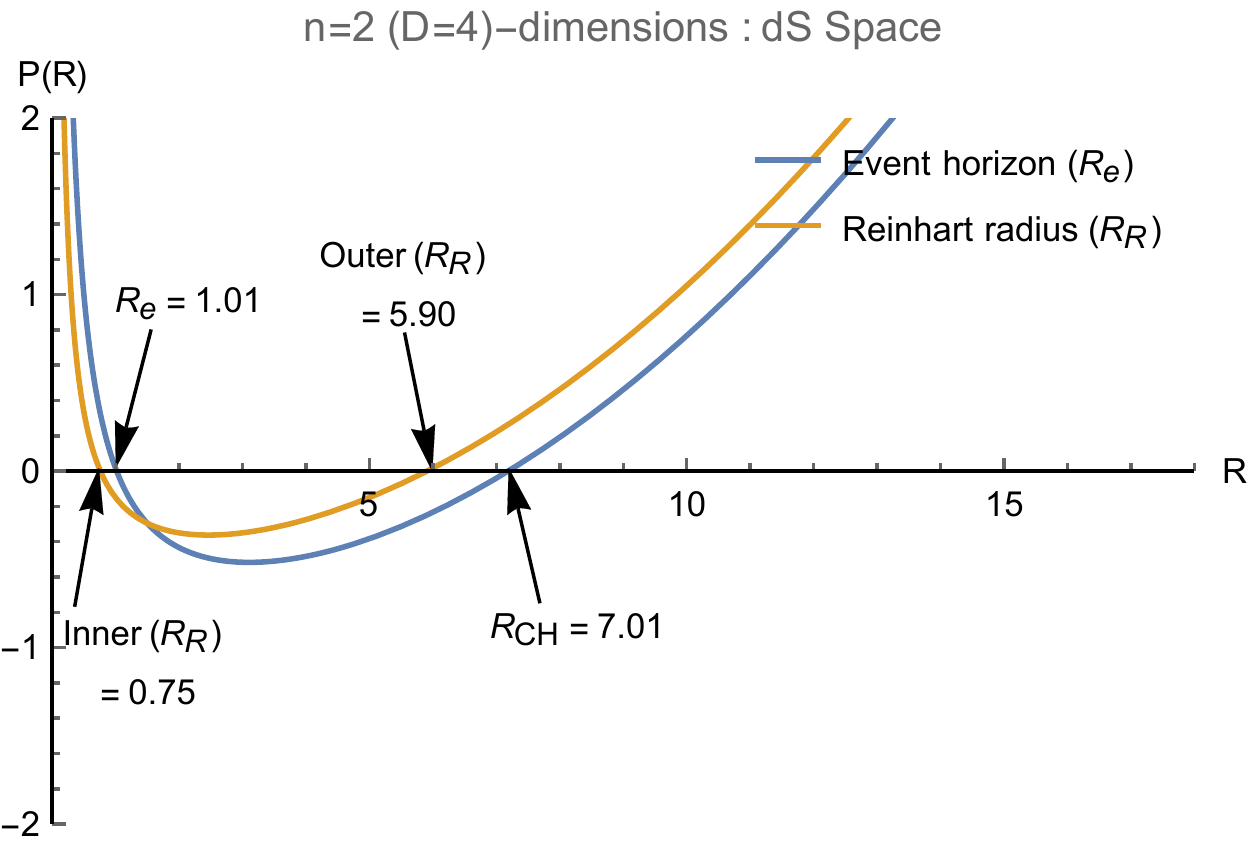} }
  \subfigure[]{ \includegraphics[width=0.31\textwidth,clip]{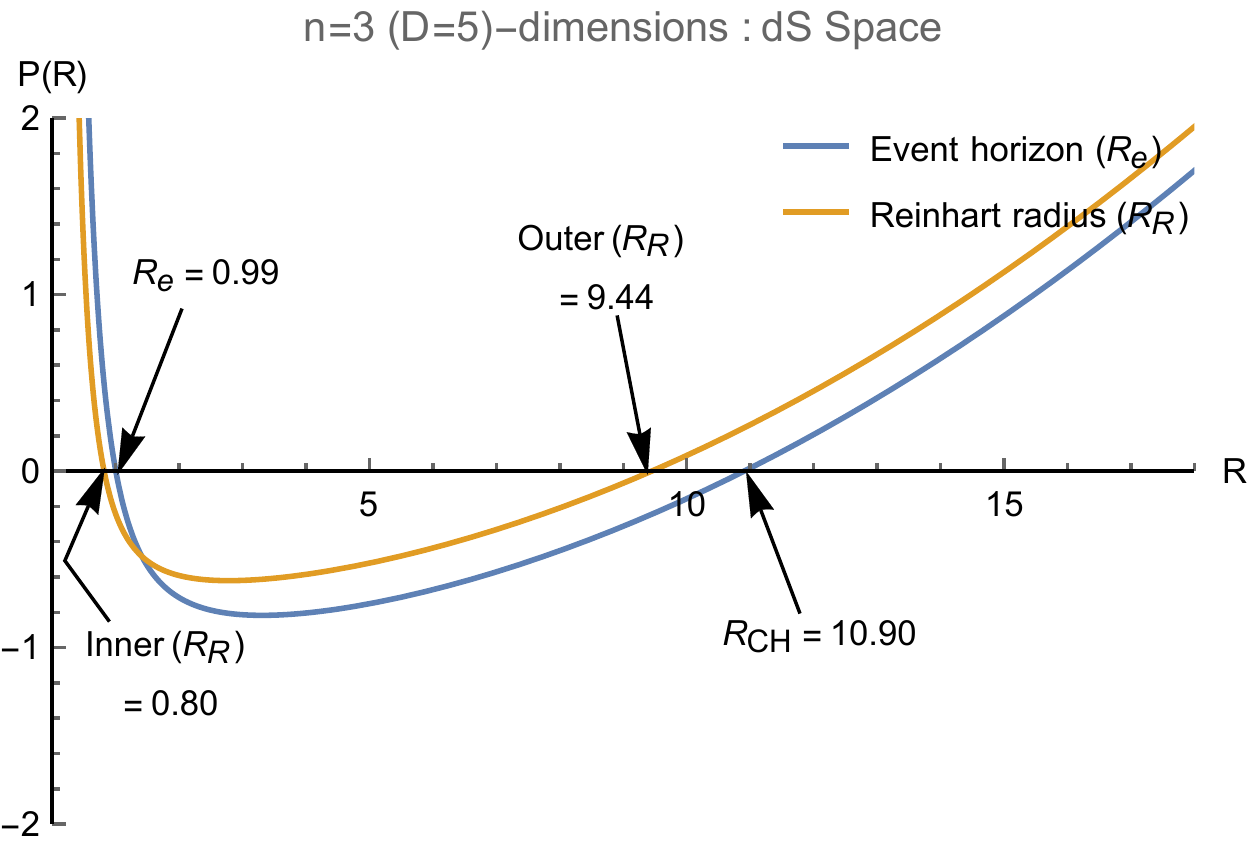} }
  \subfigure[]{ \includegraphics[width=0.31\textwidth,clip]{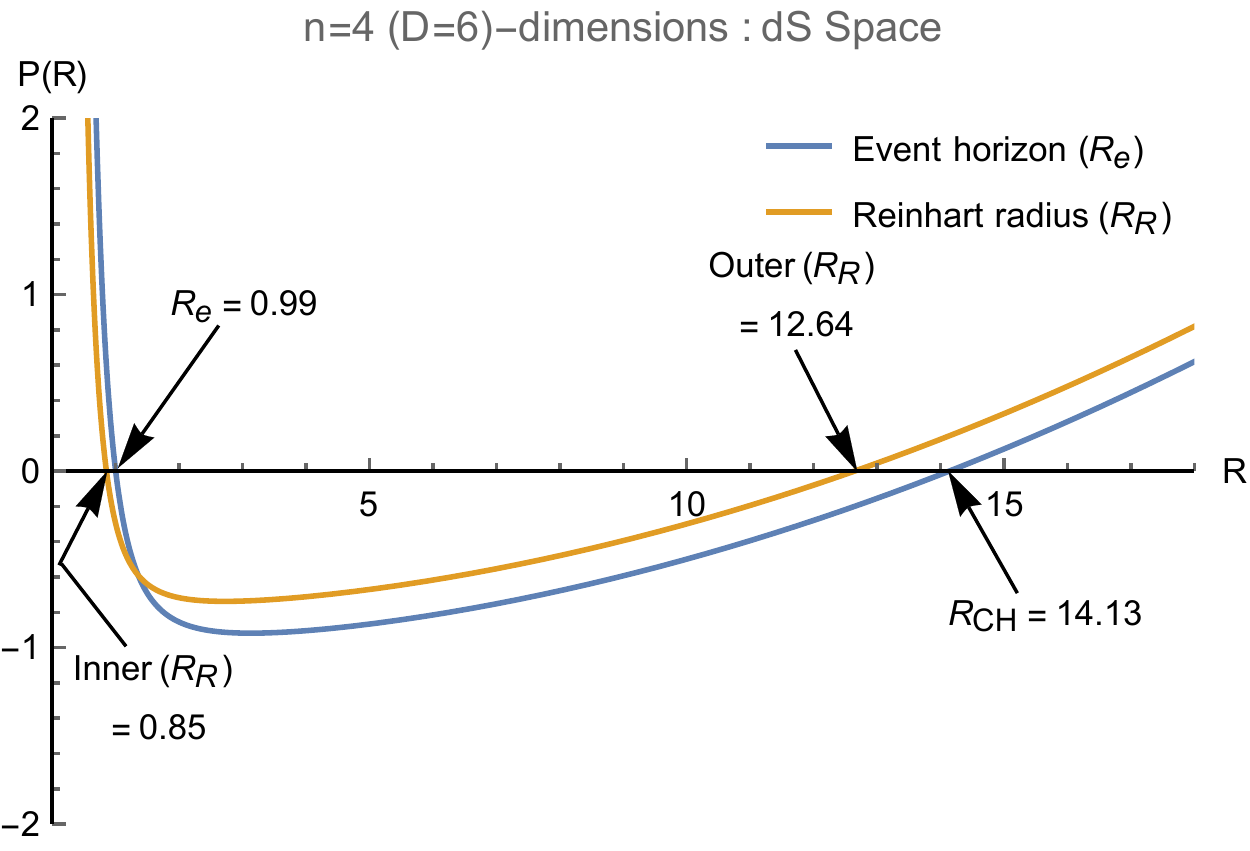} }
  \subfigure[]{ \includegraphics[width=0.31\textwidth,clip]{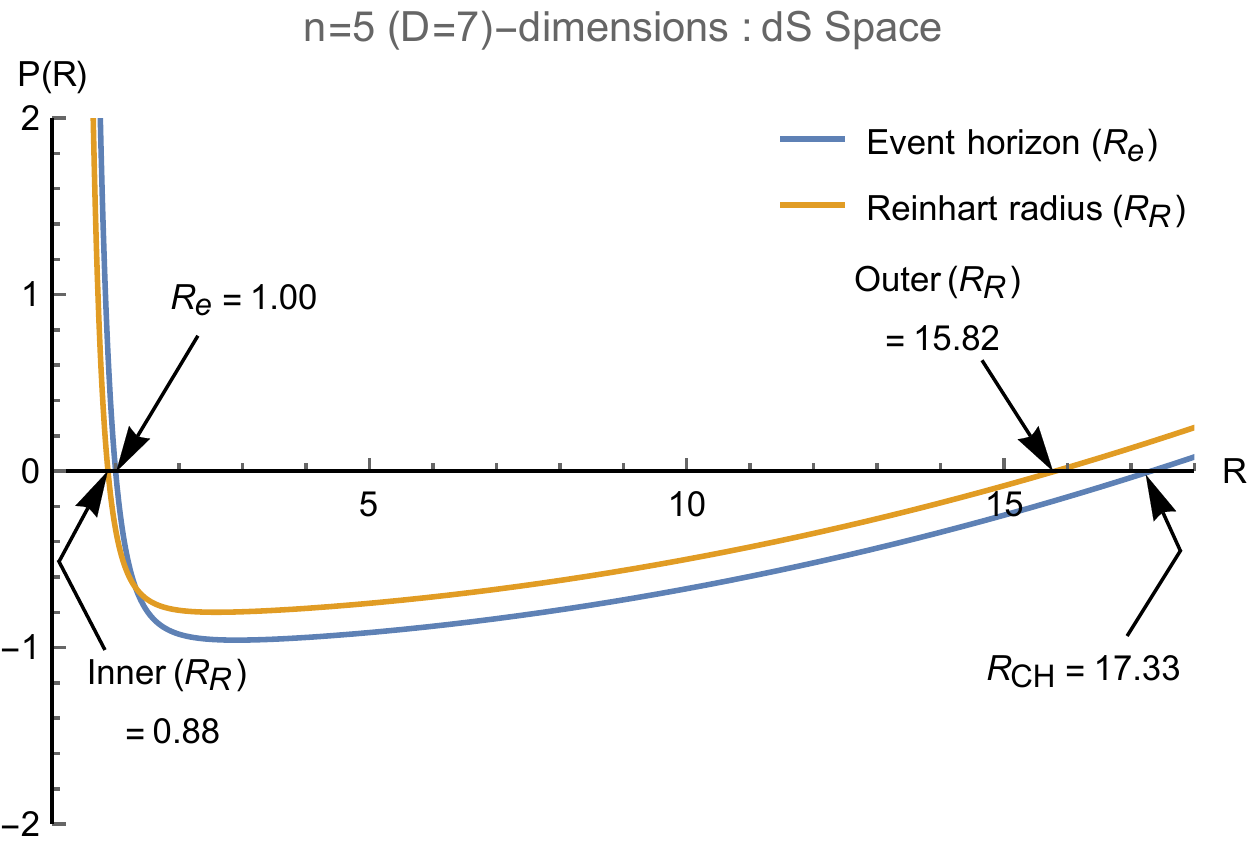}}
\caption{\label{fig:3} {The graph (a) shows the location of all the event horizons $R_e$ and cosmological horizons $R_{CH}$ corresponding to the polynomial $P(R = R_e)$ of Eq.(\ref{eqn53}). The graph (b) shows the location of all the inner and outer Reinhart radius $R_{R}$ corresponding to the polynomial $P(R=R_{R})$ of Eq.(\ref{eqn54}) for D = 4, 5, 6, and 7 dimensions in de Sitter (dS) spacetime. The graphs (d), (e), and (f) show that the inner $R_{R}$ always lies inside the $R_e$ and outer $R_{R}$ lies between $R_e$ and $R_{CH}$. Here we take the cosmological constant $\Lambda = 0.05$ and mass function $F(r) = 1$ for all the graphs}.}
\end{figure*}

If the mass function $F(r)=const.$, then the change in mass function of the dust cloud vanishes, i.e., $F'(r) = 0$, and the event horizon will be static. Let cosmological constant $\Lambda \neq 0$, then from Eq.(\ref{eqn46}), we get
\begin{multline}\label{eqn52}
    \bigg(\frac{2\Lambda R^2}{n(n+1)} + \frac{F(r)}{R^{n-1}} -1\bigg) \Bigg[\frac{2\Lambda R^{n+1} }{n} + \frac{(n+1)F(r)}{2} - nR^{n-1} \Bigg]\\ = 0.
\end{multline}
From Eq.(\ref{eqn52}), the solutions of the first and second brackets give event horizon $R_e$ and Reinhart radius $R_{R}$ as
\begin{equation}\label{eqn53}
     P(R = R_e) =\frac{2\Lambda R^2_{e}}{n(n+1)} + \frac{F}{R^{n-1}_{e}} -1 = 0
\end{equation}
and
\begin{equation}\label{eqn54}
     P(R = R_{R}) = \frac{2\Lambda R^2_{R}}{n(n+1)} + \frac{F}{2R^{n-1}_{R}} - \frac{n}{n+1} =0.
\end{equation}
Both of the Eqs. (\ref{eqn53}) and (\ref{eqn54}) are $(n+1)$-dimensional polynomials and finding analytical roots is in general difficult. We obtain good insight into the relative locations of the event horizons $R_e$ and the $R_{R}$ by plotting the graphs of the polynomials explicitly. We observe that $R_{R}$ usually lies in the interior of a black hole. In the presence of a positive cosmological constant, we get another root for $R_{R}$, which is at a radius smaller than the cosmological horizon $R_{CH}$. These facts are illustrated in the plots given in the Figs. \ref{fig:2} and \ref{fig:3}.
\subsection{D-dimensional cosmological scenario}
We now explore the presence and evolution of the Reinhart radius for the cosmological solutions in D-dimensions. We can think of the cosmological solutions as being valid for the entire Universe or we can think of the solutions as a homogeneous interior of an evolving star. If we assume the latter, we need to define an outer comoving shell $r_0$ beyond which there is the vacuum.  The mass function for homogeneous dust in D$(= n+2)$-dimension is defined as $F(r) = \frac{2g}{n(n+1)}r^{n+1}$, where $g=k\epsilon(0,r) >0$ and $R = R(t,r) = r a(t)$. The dynamics of various cases are given in the Appendix. Substituting these values in Eq.(\ref{eqn46}), we get
\begin{widetext}
\begin{equation}\label{eqn60}
    \bigg(\frac{4(\Lambda a^2 + ga^{1-n})^2}{n^3 (n+1)}\bigg) r^{4} - \bigg(\frac{2\Lambda a^2(2n+1) + g(3n+1)a^{1-n}}{n^2(n+1)}\bigg) r^{2} + 1 = 0.
\end{equation}
\end{widetext}
In case (1) when $\Lambda = 0$, i.e., without cosmological constant, the solution of Eq.(\ref{eqn60}) gives Reinhart radius $r_{R}$ as
\begin{widetext}
\begin{equation}\label{eqn71}
    r_{R} = \frac{\sqrt{n}}{2}\Bigg(\frac{(3n+1) \pm \sqrt{(3n+1)^2 - 16n(n+1)}}{2ga^{1-n}}\Bigg)^{1/2} =
      \begin{cases}
         0 &; \ if \ n = 0 \ i.e. \ (\ 2-dimensions)\\
         Im &; \ if \ n\geq 1 \ i.e. \ (\ \geq3-dimensions).
    \end{cases}
\end{equation}
\end{widetext}
Equation (\ref{eqn71}) shows that the Reinhart radius does not exist in any number of dimensions when the cosmological constant is zero.\\
In case(2) for a general case ( when $\Lambda \neq 0$), the solution of Eq.(\ref{eqn60}) gives Reinhart radius $r_{R}$ as
\begin{widetext}
\begin{equation}\label{eqn61}
    r_{R} = \frac{\sqrt{n}}{2}\Bigg(\frac{[2\Lambda a^2(2n+1) + g(3n+1)a^{1-n}] \pm \sqrt{[2\Lambda a^2(2n+1) + g(3n+1)a^{1-n}]^2-16n(n+1)(\Lambda a^2 + ga^{1-n})^2}}{2(\Lambda a^2 + ga^{1-n})^2}\Bigg)^{1/2}.
\end{equation}
\end{widetext}
There are two possible situations that arise from Eq.(\ref{eqn61}) for the validity of Reinhart radius and which we discuss as follows:\\
(a). When $\Lambda <0$, i.e., anti-de Sitter spacetime, we replace the cosmological constant $\Lambda$ with $-\Lambda$ in Eq.(\ref{eqn61}), and we get the Reinhart radius $r_{R}$ as
\begin{widetext}
\begin{equation}\label{eqn61'}
    r_{R} = \frac{\sqrt{n}}{2}\Bigg(\frac{[-2|\Lambda| a^2(2n+1) + g(3n+1)a^{1-n}] \pm \sqrt{[-2|\Lambda| a^2(2n+1) + g(3n+1)a^{1-n}]^2-16n(n+1)(|\Lambda| a^2 + ga^{1-n})^2}}{2(-|\Lambda| a^2 + ga^{1-n})^2}\Bigg)^{1/2}.
\end{equation}
\end{widetext}
Now, let us define $x =[-2|\Lambda| a^2(2n+1) + g(3n+1)a^{1-n}]$, $y = 4\sqrt{n(n+1)}(-|\Lambda| a^2 + ga^{1-n})$, and $z = \sqrt{2}(-|\Lambda| a^2 + ga^{1-n})$. There are two possible cases that arise on $x$.\\
(i). Suppose $x$ is negative, i.e., $x<0$, then Eq.(\ref{eqn61'}) gives Reinhart radius $r_{R}$ as follows:
\begin{equation}\label{eqn71'}
    r_{R} = \frac{\sqrt{n}}{2}\bigg(\frac{-|x| \pm \sqrt{x^2-y^2}}{z^2}\bigg).
\end{equation}
Now we check the existence of Reinhart radius $r_{R}$ defined in Eq.(\ref{eqn71'}) as follows
\begin{equation}\label{eqn72'}
    \begin{split}
        x^2 > x^2 - y^2\\
     \Rightarrow |x| > \sqrt{x^2-y^2}\\
      \Rightarrow -|x| < - \sqrt{x^2-y^2}\\
      \Rightarrow -|x| + \sqrt{x^2-y^2} <0\\
      \ and \ also,\ -|x| - \sqrt{x^2-y^2} <0.
    \end{split}
\end{equation}
Equation (\ref{eqn72'}) shows that the terms $-|x|\pm \sqrt{x^2-y^2}<0$, which means $r_{R}$ becomes imaginary and hence doesn't exist. \\
(ii). Suppose $x$ is positive, i.e., $x>0$, then Eq.(\ref{eqn61'}) gives Reinhart radius $r_{R}$ as
\begin{equation}\label{eqn73'}
    r_{R} = \frac{\sqrt{n}}{2}\bigg(\frac{x \pm \sqrt{x^2-y^2}}{z^2}\bigg).
\end{equation}
Now, if $x>0$, then we get a condition on scale parameter $a(t)$ as $a^{n+1}<g(3n+1)/2|\Lambda| (2n+1)$. Using this condition for any value of $n$ (i.e., $n=1,2,3,4,...$) the term $\sqrt{x^2-y^2}$ in Eq.(\ref{eqn73'}) becomes imaginary. Hence, $r_{R}$ does not exist for positive $x$ also. Therefore, these two  cases prove that the Reinhart radius $r_{R}$ does not exist in D-dimensional cosmological AdS spacetime.\\
(b). When $\Lambda >0$, i.e., de Sitter spacetime, from the Eq.(\ref{eqn61}) the Reinhart radius $r_{R}$ is positive only when the terms inside the square root are positive, i.e.,
\begin{multline}\label{eqn63}
    [2\Lambda a^2(2n+1) + g(3n+1)a^{1-n}]^2\\ -16n(n+1)(\Lambda a^2 + ga^{1-n})^2 >0.
\end{multline}
The inequality in Eq.(\ref{eqn63}) gives the condition on the scale parameter $a(t)$ as
\begin{equation}\label{eqn66}
    \begin{split}
     a^{n+1} > -\frac{g}{2\Lambda}\Bigg[\frac{4\sqrt{n(n+1)}+(3n+1)}{(2n+1)+2\sqrt{n(n+1)}}\Bigg]\\ \ and \ a^{n+1} > \frac{g}{2\Lambda}\Bigg[\frac{4\sqrt{n(n+1)}-(3n+1)}{(2n+1)-2\sqrt{n(n+1)}}\Bigg].
    \end{split}
\end{equation}
From the inequalities in Eq.(\ref{eqn66}), the general condition on scale parameter $a(t)$ in de Sitter spacetime is
\begin{equation}\label{eqn67}
    a^{n+1}> \frac{g}{2\Lambda}\Bigg[\frac{4\sqrt{n(n+1)}-(3n+1)}{(2n+1)-2\sqrt{n(n+1)}}\Bigg],
\end{equation}
where $4\sqrt{n(n+1)} > 3n+1$ and $2n+1 > 2\sqrt{n(n+1)}$, which shows that the terms in the square bracket of Eq.(\ref{eqn67}) are positive. Hence, in the above inequality, the scale parameter $a(t)$ is positive, which means there exists Reinhart radius $r_{R}$ in de Sitter spacetime cosmological solution in D-dimensions.
\section{Estimation of the Volume of evolving black hole in D-dimensions}
In the cases discussed in the previous section, we note the following. The nonexistence of Reinhart radius is shown for the case of homogeneous and isotropic scenarios like FLRW spacetime or Oppenheimer-Snyder collapse. In other generic circumstances, the Reinhart radius can form. In a realistic scenario, the dust density reduces continuously to zero as one moves outwards. In such models, the Reinhart radius does form and can evolve continuously. The evolving Reinhart radius finally asymptotes to the Reinhart radius of the Schwarzschild black hole (or other relevant exterior spacetime).  The result in the paper implies that if we take the Oppenheimer-Snyder collapsing dust, then the Reinhart radius is never inside the dust cloud but can be there in the exterior part of the spacetime. In the cases where the Reinhart radius does not exist, a closed-form expression along the lines of the work done in \cite{CR} is not possible and the volume estimation has to be done numerically.\par
 Based on the work done in the article and also the work in \cite{CL}, the collapsing matter solutions come in two categories in the context of the article. In one category, like the cases discussed in \cite{CL}, the Reinhart radius exists within the interior solution, in which case, the maximal volume can be estimated using the Reinhart radius. In this case, one can estimate the volume of the black hole using the Reinhart radius. This is presented below right after the discussion of the second case. \par
In the second case, the Reinhart radius is not present in the interior of the dust cloud (for example Oppenheimer-Snyder dust collapse or cosmological solution with negative cosmological constant as shown in the article). In such cases, the volume in the interior of the cloud cannot be simplified using the Reinhart radius but, as pointed out earlier, needs to be numerically evaluated. This also implies that the Reinhart radius is present in the region outside of the matter cloud (usually within the event horizon). The volume in the exterior part, though,  can then be estimated using the Reinhart radius.\par
We now show how one can estimate the volume of collapsing matter scenarios that have Reinhart radius using a closed form expression. An example we consider is a collapsing scenario consisting of shell-by-shell collapse of matter without cosmological constant in D ($=n+2$ ) dimensions. In such a scenario, the Reinhart radius undergoes a series of jumps given by $R_R = \big(\frac{n+1}{2n}F(r_i)\big)^{\frac{1}{n-1}}$, where $F(r_i)$ is the Misner-Sharp mass of the black hole formed by the collapse of the first $i$ shells. As is easily seen, the Reinhart is a monotonically increasing function of the Misner-Sharp mass. This fact can be used to set a lower bound on the volume of the black hole. We show this below. The metric between the shells is Schwarzschild and is defined as
\begin{multline}
    ds^2 = -\bigg(1-\frac{F(r_i)}{R^{n-1}}-\frac{2\Lambda R^2}{n(n+1)}\bigg)dt^2\\ + \bigg(1-\frac{F(r_i)}{R^{n-1}}-\frac{2\Lambda R^2}{n(n+1)}\bigg)^{-1}dR^2 + R^2d\Omega^2_n.
    \label{exteriormetric1}
\end{multline}
We first note that the interior of the matter cloud can be matched to the Eddington-Finkelstein spacetime. The mass function $F(r_i)$ gets related to the combined ADM mass $M_i$ of the first $i$th shells using the Eq. (\ref{9}). In the Eddington-Finkelstein coordinates $(v, R,\theta, \phi)$ the metric (\ref{exteriormetric1}) exterior to the shell of radius $r_i$ becomes (this metric is valid between the shells $r_i$ to $r_{i+1}$)
\begin{equation}\label{4}
    ds^2 = - N^2(r_i)dv^2 + 2dvdR + R^2d\Omega^2_n,
\end{equation}
where, the lapse function $N^2(r_i) =\bigg(1-\frac{F(r_i)}{R^{n-1}}-\frac{2\Lambda R^2}{n(n+1)}\bigg)$. For $R=const.$ hypersurface the volume of the metric (\ref{4}) is defined as
\begin{multline}\label{5}
    V^{(D)} = \int \sqrt{-g}dv d\Omega_n \\ 
    = \int \sqrt{R^{2n}\bigg(1-\frac{F(r_i)}{R^{n-1}}-\frac{2\Lambda R^2}{n(n+1)}\bigg)}dv \int d\Omega_n \\= \frac{2\pi^{\frac{n+1}{2}}}{\Gamma(\frac{n+1}{2})}\int \sqrt{R^{2n}\bigg(1-\frac{F(r_i)}{R^{n-1}}-\frac{2\Lambda R^2}{n(n+1)}\bigg)} dv.
\end{multline}
 Now, from the Eqs. (\ref{9}) and (\ref{5}) we can calculate the volume of black holes using the Reinhart radius after the shell of radius $r_i$ has collapsed.\par 
 We estimate the volume for the easier case of $\Lambda=0$ and D(=n+2)-dimensional Schwarzschild black hole. Here Reinhart radius is obtained as $R_R = \big(\frac{n+1}{2n}F(r_i)\big)^{\frac{1}{n-1}}$, where $F(r_i)$ is given in Eq.(\ref{9}). The volume of the black hole after the shell with label $i$ has collapsed is given from Eq.(\ref{5}) as
\begin{multline}\label{9'}
    V^{(D)} = \frac{2\pi^{\frac{n+1}{2}}}{\Gamma(\frac{n+1}{2})}\int \sqrt{R^{2n}_R\bigg(1-\frac{F(r_i)}{R^{n-1}_R}\bigg)}dv \\
    = \frac{2\pi^{\frac{n+1}{2}}}{\Gamma(\frac{n+1}{2})}\bigg(\frac{n+1}{2n}F(r_i)\bigg)^{\frac{n}{n-1}}\bigg(\bigg|1-\frac{2n}{n+1}\bigg|\bigg)^{1/2}v
\end{multline}
Now the Misner-sharp mass $F(r)$ is a monotonically increasing function of $r$ (unless we consider the $r_f$, beyond which there is no further matter that will collapse, in which case, $F(r)$ will be a constant or the case of Hawking radiation for which $F(r)$ decreases with time). The Reinhart radius too is a monotonically increasing function of $r$ as more shells collapse. Therefore, the asymptotic volume of the black hole is always greater than or equal to (when the shell is the outermost one) the volume found in Eq. (\ref{9'}).  So during the collapse, if we find that there is a Reinhart radius available for a shell of radius $r_i$, then the eventual asymptotic volume is greater than or equal to that of Eq. (\ref{9'}). So we can use the Reinhart radius to get lower bounds on the asymptotic volume of the black hole formed during the collapse. 
\section{Kodama Vector for spacelike hypersurfaces in spherically symmetric spacetime}
We now show that the Kodama vector is tangential to the maximal hypersurface at these Reinhart radii. The trace of the extrinsic curvature vanishes at Reinhart radius by definition \cite{Reinhart,FE}. Kodama vector in a spherically symmetric spacetime is defined in \cite{HK,VF}. We consider the metric for the D$(= n+2)$-dimensional spherically symmetric dust cloud, defined as
\begin{equation}\label{eqn82}
    ds^2 = -dt^2 + R'^2(t, r) dr^2 + R^2(t,r) d{\Omega}^2_{n}.
\end{equation}
The 2-metric in the coordinate chart $(t,r)$ is given by
\begin{equation}\label{eqn83}
    ds^2_{2} = -dt^2+R'^2dr^2.
\end{equation}
The two-dimensional volume form in $(t,r)$ coordinates is expressed as
\begin{equation}\label{eqn84}
    \epsilon=R'dt\wedge dr.
\end{equation}
Using the standard definition of Kodama vector \cite{HK,VF}, $K^a=\epsilon^{ab}\partial_{b}R$, where $(a, b = t, r)$, $R(t,r)$ is the area radius, $\epsilon^{ab}$ is the volume form of the 2-metric of Eq.(\ref{eqn83}), and we evaluate the components to be
\begin{equation}
    K^t= -1 \  \ and \  \ K^r= \frac{\dot{R}}{R'}.
\end{equation}
Now, evaluating the dot product with the normal vector $n_{\alpha} = (n_t, n_r) = (\dot R, R')$ obtained from Eq.(\ref{constR17}), we find that 
\begin{equation}
    K^{\alpha}n_{\alpha} = K^t n_t + K^r n_r = (-1)\times \dot R + \bigg(\frac{\dot R}{R'}\bigg)\times R' = 0.
\end{equation}
This shows that the Kodama vector is tangential to the maximal hypersurface at the Reinhart radius. We note that the above result is independent of whether the Kodama vector is spacelike or timelike. In fact for the cases that were discovered in \cite{CR,BJ,Ong,YCO,CL,XY}, the Kodama vector is spacelike and is tangential to the maximal hypersurface at the Reinhart radius. Another interesting observation is that at the Reinhart radius, both the normal vector to the hypersurface and the Kodama vector have vanishing divergence.
\section{Conclusions}
In this work, we address a few aspects concerning the maximal hypersurface of a black hole in a dynamically evolving scenario. We considered the spherically symmetric collapse of dust clouds, generalized to D-dimensions since the model is analytically tractable. We have carried out the analysis separately for $(2+1)$-dimensions and grouped the other dimensions together. This is due to the fact that $(2+1)$-dimensional gravity is fundamentally different from other dimensions owing to the topological nature of gravity in $(2+1)$-dimensions. The dimensions $D>3$ are qualitatively similar to each other. For the evolving setting, we choose the Lemaitre-Tolman-Bondi model generalized to D-dimensions since the model has simplicity in terms of analytical expressions while capturing the core essence of the problem.\par 
We obtain the differential equation for the maximal hypersurface using the variational technique developed in \cite{CR}. We set up a Lagrangian whose solution to the Euler-Lagrange equation yields the maximal hypersurface in an evolving setting. By choosing the appropriate boundary values for the solutions one can arrive at the maximal volume inside a trapped region which is in the process of evolving. The same procedure is generalized to D-dimensions. We present the equations by considering a subclass of Lemaitre-Tolman-Bondi models, the homogeneous dust evolution where the expressions greatly simplify.\par
We analyze an interesting region of the maximal hypersurfaces, which we denote as ``Reinhart radius" ($R_{R}$). The reason for this nomenclature is due to the role these points play in the estimation of the maximal volume inside a black hole. Identified first by Reinhart \cite{Reinhart} 1973, these Reinhart points were found in various other black holes. In this article, we explored the existence and evolution of these points during the course of the formation of black holes. We have identified an interesting property of these points in relation to the maximal hypersurfaces. These points are located where the Kodama vector becomes tangential to the maximal hypersurface. The geometrical meaning and consequence of this observation are left for future considerations. The Kodama vector works as a substitute for a timelike Killing vector in scenarios that do not have a timelike Killing vector. Kodama vector, when it is timelike, has been used to define surface gravity in a dynamically evolving setting. In this article, we find another role of the Kodama vector, viz., it is used to pinpoint the Reinhart radii of a maximal hypersurface. We note that inside the black holes, the Kodama vector is spacelike.\par
We develop a formula to find the location of $R_{R}$ in terms of coordinate invariants like area radius, cosmological constants, the principle value of the energy-momentum tensor, and Misner-Sharp mass. Using the formula one can locate the Reinhart radius in various situations. We have explicitly evaluated the location of $R_{R}$ for the vacuum case and black hole case with and without the cosmological constant. We have presented our analysis and compared the $R_{R}$ with the position of the event horizon and cosmological horizon. We showed that in the black hole scenario, the $R_{R}$ is located within the event horizon. If there is a positive cosmological constant, then we showed that $R_{R}$ lies at an area radius smaller than the cosmological horizon. When we consider an evolving situation,  we use the collapse of a homogeneous dust cloud. This can be viewed as a cosmological solution or the collapse of a star with a homogeneous distribution of dust. We showed that for the case of the Oppenheimer-Snyder scenario and the collapse with a negative cosmological horizon, there is no real solution for $R_{R}$, and therefore, $R_{R}$ does not exist. For the dust evolution in the presence of a positive cosmological constant, we showed that $R_{R}$ exists provided the evolving scale factor crosses a certain critical value. We show that during a collapsing scenario, we can use the Reinhart radius, whenever available, to get lower bounds on the asymptotic volume of the black hole formed during the collapse.\par 
The analysis raises many questions that are left for future consideration. Does the relation between the Kodama vector and the maximal hypersurface continue to hold in a nonspherically symmetric situation, like the Kerr family of solutions? A timelike Kodama vector has been used to define various quantities that have thermodynamic interpretation like surface gravity, etc. in dynamical situations. Does a spacelike Kodama vector also have a geometric interpretation? The Lagrangian formulation for the maximal hypersurface in Kerr/Kerr-AdS/Kerr-Newman/Kerr-De Sitter has not been formulated though there are many interesting papers estimating the volume of the interior of the Kerr family of black holes \cite{BJ, XY}. We also note that the radius used to estimate the maximum volume in these papers does not obey the property of the trace of extrinsic curvature vanishing and hence is not Reinhart radii. The Lagrangian formulation of the Kerr family is a work in progress.\par
We note that at $R_{R}$ both the normal vector and its tangent have zero divergence. Is there a special geometric meaning associated with $R_{R}$ owing to the above property?  These questions are left open.
\begin{acknowledgments}
     We would like to thank our institute BITS Pilani Hyderabad campus for providing the required infrastructure to carry out this research work. S. M. would further like to thank the funding agency, Council of Scientific and Industrial Research (CSIR), Government of India, File No. 09/1026(11329)/2021-EMR-I, for providing the necessary fellowship to support this research work.
\end{acknowledgments}
\appendix*
\section{SOLUTION OF THE SCALE PARAMETER \textit{a(t)} FOR THE HOMOGENEOUS DUST EVOLUTION}
As we know the area radius $R(t,r)$ and mass function $F(r)$ of homogeneous dust are defined as
\begin{equation}\label{homogeneous23}
    R(t,r) = r a(t) \ and \ F(r) = \frac{2g}{n(n+1)}r^{n+1},
\end{equation}
and also 
\begin{equation}\label{rdotsquare24}
\dot R^2 = \frac{2\Lambda}{n(n+1)}R^2 + \frac{2g}{n(n+1)R^{n-1}}r^{n+1}. 
\end{equation}
From Eqs.(\ref{homogeneous23}) and (\ref{rdotsquare24}) we get
\begin{equation}\label{eqn25}
    [\dot{a}(t)]^2 = \frac{2\Lambda}{n(n+1)}[a(t)]^2 + \frac{2g}{n(n+1)[a(t)]^{n-1}},
\end{equation}
and the solution of Eq.(\ref{eqn25}) gives the scale parameter $a(t)$ for different regions of spacetime based on cosmological constant $(\Lambda)$.
\subsection{\textbf{For zero cosmological constant}}
For zero cosmological constant $\Lambda = 0$, Eq.(\ref{eqn25}) becomes
\begin{equation}\label{eqn26}
    [\dot{a}(t)]^2 = \frac{2g}{n(n+1)[a(t)]^{n-1}} 
    \Rightarrow \frac{da(t)}{dt} =\pm \frac{1}{a^{\frac{n-1}{2}}}\sqrt{\frac{2g}{n(n+1)}}.
\end{equation}
and the solution of Eq.(\ref{eqn26}) with initial scale parameter $a(0) = 1$ gives
\begin{equation}
    a(t) = \Bigg(1 + \sqrt{\frac{g(n+1)}{2n}}t\Bigg)^{\frac{n+1}{2}} \ and \  \Bigg(1 - \sqrt{\frac{g(n+1)}{2n}}t\Bigg)^{\frac{n+1}{2}}. 
\end{equation}

\subsection{\textbf{For de Sitter spacetime}}
For positive cosmological constant $(\Lambda > 0)$, the change in scale parameter is defined as
\begin{equation}\label{lambdapositive28}
    \begin{split}
     [\dot{a}(t)]^2 = \frac{2\Lambda}{n(n+1)}[a(t)]^2 + \frac{2g}{n(n+1)[a(t)]^{n-1}}\\
     \Rightarrow \frac{da(t)}{dt} =\pm \sqrt{\frac{2\Lambda}{n(n+1)}[a(t)]^2 + \frac{2g}{n(n+1)[a(t)]^{n-1}}}.
     \end{split}
\end{equation}
The solutions of Eq.(\ref{lambdapositive28}) with initial condition $a(0) = 1$ are
\begin{multline}
        a(t) = \Bigg[\sqrt{\frac{g}{\Lambda}}sinh\Bigg(\sqrt{\frac{\Lambda (n+1)}{2n}}t + arcsinh\sqrt{\frac{\Lambda}{g}}\Bigg)\Bigg]^{\frac{2}{n+1}}\\ \ and \ 
        \Bigg[-\sqrt{\frac{g}{\Lambda}}sinh\Bigg(\sqrt{\frac{\Lambda (n+1)}{2n}}t - arcsinh\sqrt{\frac{\Lambda}{g}}\Bigg)\Bigg]^{\frac{2}{n+1}}.
\end{multline}
\subsection{\textbf{For anti-de Sitter spacetime}}
For negative cosmological constant $(\Lambda < 0)$, the change in scale parameter is defined as
\begin{equation}\label{lambdanegative30}
    \begin{split}
     [\dot{a}(t)]^2 = -\frac{2\Lambda}{n(n+1)}[a(t)]^2 + \frac{2g}{n(n+1)[a(t)]^{n-1}}\\
     \Rightarrow \frac{da(t)}{dt} =\pm \sqrt{-\frac{2\Lambda}{n(n+1)}[a(t)]^2 + \frac{2g}{n(n+1)[a(t)]^{n-1}}}.
     \end{split}
\end{equation}
The solutions of Eq.(\ref{lambdanegative30}) with initial condition $a(0) = 1$ are
\begin{multline}
        a(t) = \Bigg[\sqrt{\frac{g}{\Lambda}}sin\Bigg(\sqrt{\frac{\Lambda (n+1)}{2n}}t + arcsin\sqrt{\frac{\Lambda}{g}}\Bigg)\Bigg]^{\frac{2}{n+1}}\\ \ and \ 
        \Bigg[-\sqrt{\frac{g}{\Lambda}}sin\Bigg(\sqrt{\frac{\Lambda (n+1)}{2n}}t - arcsin\sqrt{\frac{\Lambda}{g}}\Bigg)\Bigg]^{\frac{2}{n+1}}.
\end{multline}
\\
\bibliography{bibitems}

\begin{thebibliography}{20}%
\makeatletter
\providecommand \@ifxundefined [1]{%
 \@ifx{#1\undefined}
}%
\providecommand \@ifnum [1]{%
 \ifnum #1\expandafter \@firstoftwo
 \else \expandafter \@secondoftwo
 \fi
}%
\providecommand \@ifx [1]{%
 \ifx #1\expandafter \@firstoftwo
 \else \expandafter \@secondoftwo
 \fi
}%
\providecommand \natexlab [1]{#1}%
\providecommand \enquote  [1]{``#1''}%
\providecommand \bibnamefont  [1]{#1}%
\providecommand \bibfnamefont [1]{#1}%
\providecommand \citenamefont [1]{#1}%
\providecommand \href@noop [0]{\@secondoftwo}%
\providecommand \href [0]{\begingroup \@sanitize@url \@href}%
\providecommand \@href[1]{\@@startlink{#1}\@@href}%
\providecommand \@@href[1]{\endgroup#1\@@endlink}%
\providecommand \@sanitize@url [0]{\catcode `\\12\catcode `\$12\catcode
  `\&12\catcode `\#12\catcode `\^12\catcode `\_12\catcode `\%12\relax}%
\providecommand \@@startlink[1]{}%
\providecommand \@@endlink[0]{}%
\providecommand \url  [0]{\begingroup\@sanitize@url \@url }%
\providecommand \@url [1]{\endgroup\@href {#1}{\urlprefix }}%
\providecommand \urlprefix  [0]{URL }%
\providecommand \Eprint [0]{\href }%
\providecommand \doibase [0]{https://doi.org/}%
\providecommand \selectlanguage [0]{\@gobble}%
\providecommand \bibinfo  [0]{\@secondoftwo}%
\providecommand \bibfield  [0]{\@secondoftwo}%
\providecommand \translation [1]{[#1]}%
\providecommand \BibitemOpen [0]{}%
\providecommand \bibitemStop [0]{}%
\providecommand \bibitemNoStop [0]{.\EOS\space}%
\providecommand \EOS [0]{\spacefactor3000\relax}%
\providecommand \BibitemShut  [1]{\csname bibitem#1\endcsname}%
\let\auto@bib@innerbib\@empty
\bibitem [{\citenamefont {Parikh}(2006)}]{MP}%
  \BibitemOpen
  \bibfield  {author} {\bibinfo {author} {\bibfnamefont {M.~K.}\ \bibnamefont
  {Parikh}},\ }\bibfield  {title} {\bibinfo {title} {Volume of black holes},\
  }\href {https://doi.org/10.1103/PhysRevD.73.124021} {\bibfield  {journal}
  {\bibinfo  {journal} {Phys. Rev. D}\ }\textbf {\bibinfo {volume} {73}},\
  \bibinfo {pages} {124021} (\bibinfo {year} {2006})}\BibitemShut {NoStop}%
\bibitem [{\citenamefont {Cveti\ifmmode~\check{c}\else \v{c}\fi{}}\ \emph
  {et~al.}(2011)\citenamefont {Cveti\ifmmode~\check{c}\else \v{c}\fi{}},
  \citenamefont {Gibbons}, \citenamefont {Kubiz\ifmmode~\check{n}\else
  \v{n}\fi{}\'ak},\ and\ \citenamefont {Pope}}]{Gibbons}%
  \BibitemOpen
  \bibfield  {author} {\bibinfo {author} {\bibfnamefont {M.}~\bibnamefont
  {Cveti\ifmmode~\check{c}\else \v{c}\fi{}}}, \bibinfo {author} {\bibfnamefont
  {G.~W.}\ \bibnamefont {Gibbons}}, \bibinfo {author} {\bibfnamefont
  {D.}~\bibnamefont {Kubiz\ifmmode~\check{n}\else \v{n}\fi{}\'ak}},\ and\
  \bibinfo {author} {\bibfnamefont {C.~N.}\ \bibnamefont {Pope}},\ }\bibfield
  {title} {\bibinfo {title} {Black hole enthalpy and an entropy inequality for
  the thermodynamic volume},\ }\href
  {https://doi.org/10.1103/PhysRevD.84.024037} {\bibfield  {journal} {\bibinfo
  {journal} {Phys. Rev. D}\ }\textbf {\bibinfo {volume} {84}},\ \bibinfo
  {pages} {024037} (\bibinfo {year} {2011})}\BibitemShut {NoStop}%
\bibitem [{\citenamefont {Christodoulou}\ and\ \citenamefont
  {Rovelli}(2015)}]{CR}%
  \BibitemOpen
  \bibfield  {author} {\bibinfo {author} {\bibfnamefont {M.}~\bibnamefont
  {Christodoulou}}\ and\ \bibinfo {author} {\bibfnamefont {C.}~\bibnamefont
  {Rovelli}},\ }\bibfield  {title} {\bibinfo {title} {How big is a black
  hole?},\ }\href {https://doi.org/10.1103/PhysRevD.91.064046} {\bibfield
  {journal} {\bibinfo  {journal} {Phys. Rev. D}\ }\textbf {\bibinfo {volume}
  {91}},\ \bibinfo {pages} {064046} (\bibinfo {year} {2015})}\BibitemShut
  {NoStop}%
\bibitem [{\citenamefont {Bhaumik}\ and\ \citenamefont {Majhi}(2018)}]{Bibhas}%
  \BibitemOpen
  \bibfield  {author} {\bibinfo {author} {\bibfnamefont {N.}~\bibnamefont
  {Bhaumik}}\ and\ \bibinfo {author} {\bibfnamefont {B.~R.}\ \bibnamefont
  {Majhi}},\ }\bibfield  {title} {\bibinfo {title} {Interior volume of (1 +
  d)-dimensional schwarzschild black hole},\ }\href
  {https://doi.org/10.1142/S0217751X18500112} {\bibfield  {journal} {\bibinfo
  {journal} {Int. J. Mod. Phys. A}\ }\textbf {\bibinfo {volume} {33}},\
  \bibinfo {pages} {1850011} (\bibinfo {year} {2018})}\BibitemShut {NoStop}%
\bibitem [{\citenamefont {Zhang}(2015)}]{BZ}%
  \BibitemOpen
  \bibfield  {author} {\bibinfo {author} {\bibfnamefont {B.}~\bibnamefont
  {Zhang}},\ }\bibfield  {title} {\bibinfo {title} {Entropy in the interior of
  a black hole and thermodynamics},\ }\href
  {https://doi.org/10.1103/PhysRevD.92.081501} {\bibfield  {journal} {\bibinfo
  {journal} {Phys. Rev. D}\ }\textbf {\bibinfo {volume} {92}},\ \bibinfo
  {pages} {081501} (\bibinfo {year} {2015})}\BibitemShut {NoStop}%
\bibitem [{\citenamefont {Bengtsson}\ and\ \citenamefont
  {Jakobsson}(2015)}]{BJ}%
  \BibitemOpen
  \bibfield  {author} {\bibinfo {author} {\bibfnamefont {I.}~\bibnamefont
  {Bengtsson}}\ and\ \bibinfo {author} {\bibfnamefont {E.}~\bibnamefont
  {Jakobsson}},\ }\bibfield  {title} {\bibinfo {title} {Black holes: Their
  large interiors},\ }\href {https://doi.org/10.1142/S0217732315501035}
  {\bibfield  {journal} {\bibinfo  {journal} {Mod. Phys. Lett. A}\ }\textbf
  {\bibinfo {volume} {30}},\ \bibinfo {pages} {1550103} (\bibinfo {year}
  {2015})}\BibitemShut {NoStop}%
\bibitem [{\citenamefont {Ong}(2015{\natexlab{a}})}]{Ong}%
  \BibitemOpen
  \bibfield  {author} {\bibinfo {author} {\bibfnamefont {Y.~C.}\ \bibnamefont
  {Ong}},\ }\bibfield  {title} {\bibinfo {title} {Never judge a black hole by
  its area},\ }\href {https://doi.org/10.1088/1475-7516/2015/04/003} {\bibfield
   {journal} {\bibinfo  {journal} {Journal of Cosmology and Astroparticle
  Physics}\ }\textbf {\bibinfo {volume} {2015}}\bibinfo  {number} { (04)},\
  \bibinfo {pages} {003}}\BibitemShut {NoStop}%
\bibitem [{\citenamefont {Ong}(2015{\natexlab{b}})}]{YCO}%
  \BibitemOpen
\bibfield  {number} {  }\bibfield  {author} {\bibinfo {author} {\bibfnamefont
  {Y.~C.}\ \bibnamefont {Ong}},\ }\bibfield  {title} {\bibinfo {title} {The
  persistence of the large volumes in black holes},\ }\href
  {https://doi.org/10.1007/s10714-015-1929-x} {\bibfield  {journal} {\bibinfo
  {journal} {Gen. Rel. and Grav.}\ }\textbf {\bibinfo {volume} {47}},\ \bibinfo
  {pages} {88} (\bibinfo {year} {2015}{\natexlab{b}})}\BibitemShut {NoStop}%
\bibitem [{\citenamefont {Christodoulou}\ and\ \citenamefont
  {De~Lorenzo}(2016)}]{CL}%
  \BibitemOpen
  \bibfield  {author} {\bibinfo {author} {\bibfnamefont {M.}~\bibnamefont
  {Christodoulou}}\ and\ \bibinfo {author} {\bibfnamefont {T.}~\bibnamefont
  {De~Lorenzo}},\ }\bibfield  {title} {\bibinfo {title} {Volume inside old
  black holes},\ }\href {https://doi.org/10.1103/PhysRevD.94.104002} {\bibfield
   {journal} {\bibinfo  {journal} {Phys. Rev. D}\ }\textbf {\bibinfo {volume}
  {94}},\ \bibinfo {pages} {104002} (\bibinfo {year} {2016})}\BibitemShut
  {NoStop}%
\bibitem [{\citenamefont {Chew}\ and\ \citenamefont {Ong}(2020)}]{XY}%
  \BibitemOpen
  \bibfield  {author} {\bibinfo {author} {\bibfnamefont {X.~Y.}\ \bibnamefont
  {Chew}}\ and\ \bibinfo {author} {\bibfnamefont {Y.~C.}\ \bibnamefont {Ong}},\
  }\bibfield  {title} {\bibinfo {title} {Interior volume of kerr-ads black
  holes},\ }\href {https://doi.org/10.1103/PhysRevD.102.064055} {\bibfield
  {journal} {\bibinfo  {journal} {Phys. Rev. D}\ }\textbf {\bibinfo {volume}
  {102}},\ \bibinfo {pages} {064055} (\bibinfo {year} {2020})}\BibitemShut
  {NoStop}%
\bibitem [{\citenamefont {Zhang}(2019)}]{MZ}%
  \BibitemOpen
  \bibfield  {author} {\bibinfo {author} {\bibfnamefont {M.}~\bibnamefont
  {Zhang}},\ }\bibfield  {title} {\bibinfo {title} {Interior volume of
  banados-teitelboim-zanelli black hole},\ }\href
  {https://doi.org/https://doi.org/10.1016/j.physletb.2019.01.032} {\bibfield
  {journal} {\bibinfo  {journal} {Phys. Lett. B}\ }\textbf {\bibinfo {volume}
  {790}},\ \bibinfo {pages} {205} (\bibinfo {year} {2019})}\BibitemShut
  {NoStop}%
\bibitem [{\citenamefont {Maurya}\ \emph {et~al.}(2022)\citenamefont {Maurya},
  \citenamefont {Gutti},\ and\ \citenamefont {Nigam}}]{SSR}%
  \BibitemOpen
  \bibfield  {author} {\bibinfo {author} {\bibfnamefont {S.}~\bibnamefont
  {Maurya}}, \bibinfo {author} {\bibfnamefont {S.}~\bibnamefont {Gutti}},\ and\
  \bibinfo {author} {\bibfnamefont {R.}~\bibnamefont {Nigam}},\ }\bibfield
  {title} {\bibinfo {title} {Volume of a rotating black hole in 2+1
  dimensions},\ }\href
  {https://doi.org/https://doi.org/10.1016/j.physletb.2022.137381} {\bibfield
  {journal} {\bibinfo  {journal} {Phys. Lett. B}\ }\textbf {\bibinfo {volume}
  {833}},\ \bibinfo {pages} {137381} (\bibinfo {year} {2022})}\BibitemShut
  {NoStop}%
\bibitem [{\citenamefont {Reinhart}(1973)}]{Reinhart}%
  \BibitemOpen
  \bibfield  {author} {\bibinfo {author} {\bibfnamefont {B.~L.}\ \bibnamefont
  {Reinhart}},\ }\bibfield  {title} {\bibinfo {title} {Maximal foliations of
  extended schwarzschild space},\ }\href {https://doi.org/10.1063/1.1666384}
  {\bibfield  {journal} {\bibinfo  {journal} {J. Math. Phys.}\ }\textbf
  {\bibinfo {volume} {14}},\ \bibinfo {pages} {719} (\bibinfo {year}
  {1973})}\BibitemShut {NoStop}%
\bibitem [{\citenamefont {Tibrewala}\ \emph {et~al.}(2008)\citenamefont
  {Tibrewala}, \citenamefont {Gutti}, \citenamefont {Singh},\ and\
  \citenamefont {Vaz}}]{Rakesh}%
  \BibitemOpen
  \bibfield  {author} {\bibinfo {author} {\bibfnamefont {R.}~\bibnamefont
  {Tibrewala}}, \bibinfo {author} {\bibfnamefont {S.}~\bibnamefont {Gutti}},
  \bibinfo {author} {\bibfnamefont {T.~P.}\ \bibnamefont {Singh}},\ and\
  \bibinfo {author} {\bibfnamefont {C.}~\bibnamefont {Vaz}},\ }\bibfield
  {title} {\bibinfo {title} {Classical and quantum gravitational collapse in
  $d$-dimensional ads spacetime: Classical solutions},\ }\href
  {https://doi.org/10.1103/PhysRevD.77.064012} {\bibfield  {journal} {\bibinfo
  {journal} {Phys. Rev. D}\ }\textbf {\bibinfo {volume} {77}},\ \bibinfo
  {pages} {064012} (\bibinfo {year} {2008})}\BibitemShut {NoStop}%
\bibitem [{\citenamefont {Mann}\ and\ \citenamefont
  {Ross}(1993)}]{rossandmann}%
  \BibitemOpen
  \bibfield  {author} {\bibinfo {author} {\bibfnamefont {R.~B.}\ \bibnamefont
  {Mann}}\ and\ \bibinfo {author} {\bibfnamefont {S.~F.}\ \bibnamefont
  {Ross}},\ }\bibfield  {title} {\bibinfo {title} {Gravitationally collapsing
  dust in 2 + 1 dimensions},\ }\href {https://doi.org/10.1103/PhysRevD.47.3319}
  {\bibfield  {journal} {\bibinfo  {journal} {Phys. Rev. D}\ }\textbf {\bibinfo
  {volume} {47}},\ \bibinfo {pages} {3319} (\bibinfo {year}
  {1993})}\BibitemShut {NoStop}%
\bibitem [{\citenamefont {Gutti}(2005)}]{sashideep}%
  \BibitemOpen
  \bibfield  {author} {\bibinfo {author} {\bibfnamefont {S.}~\bibnamefont
  {Gutti}},\ }\bibfield  {title} {\bibinfo {title} {Gravitational collapse of
  inhomogeneous dust in (2+1) dimensions},\ }\href
  {https://doi.org/10.1088/0264-9381/22/16/007} {\bibfield  {journal} {\bibinfo
   {journal} {Class. Quant. Grav.}\ }\textbf {\bibinfo {volume} {22}},\
  \bibinfo {pages} {3223} (\bibinfo {year} {2005})}\BibitemShut {NoStop}%
\bibitem [{\citenamefont {Raviteja}\ and\ \citenamefont {Gutti}(2020)}]{KS}%
  \BibitemOpen
  \bibfield  {author} {\bibinfo {author} {\bibfnamefont {K.}~\bibnamefont
  {Raviteja}}\ and\ \bibinfo {author} {\bibfnamefont {S.}~\bibnamefont
  {Gutti}},\ }\bibfield  {title} {\bibinfo {title} {Aspects of marginally
  trapped and antitrapped surfaces in a $d$-dimensional evolving dust model},\
  }\href {https://doi.org/10.1103/PhysRevD.102.024072} {\bibfield  {journal}
  {\bibinfo  {journal} {Phys. Rev. D}\ }\textbf {\bibinfo {volume} {102}},\
  \bibinfo {pages} {024072} (\bibinfo {year} {2020})}\BibitemShut {NoStop}%
\bibitem [{\citenamefont {Estabrook}\ \emph {et~al.}(1973)\citenamefont
  {Estabrook}, \citenamefont {Wahlquist}, \citenamefont {Christensen},
  \citenamefont {DeWitt}, \citenamefont {Smarr},\ and\ \citenamefont
  {Tsiang}}]{FE}%
  \BibitemOpen
  \bibfield  {author} {\bibinfo {author} {\bibfnamefont {F.}~\bibnamefont
  {Estabrook}}, \bibinfo {author} {\bibfnamefont {H.}~\bibnamefont
  {Wahlquist}}, \bibinfo {author} {\bibfnamefont {S.}~\bibnamefont
  {Christensen}}, \bibinfo {author} {\bibfnamefont {B.}~\bibnamefont {DeWitt}},
  \bibinfo {author} {\bibfnamefont {L.}~\bibnamefont {Smarr}},\ and\ \bibinfo
  {author} {\bibfnamefont {E.}~\bibnamefont {Tsiang}},\ }\bibfield  {title}
  {\bibinfo {title} {Maximally slicing a black hole},\ }\href
  {https://doi.org/10.1103/PhysRevD.7.2814} {\bibfield  {journal} {\bibinfo
  {journal} {Phys. Rev. D}\ }\textbf {\bibinfo {volume} {7}},\ \bibinfo {pages}
  {2814} (\bibinfo {year} {1973})}\BibitemShut {NoStop}%
\bibitem [{\citenamefont {Kodama}(1980)}]{HK}%
  \BibitemOpen
  \bibfield  {author} {\bibinfo {author} {\bibfnamefont {H.}~\bibnamefont
  {Kodama}},\ }\bibfield  {title} {\bibinfo {title} {Conserved energy flux for
  the spherically symmetric system and the backreaction problem in the black
  hole evaporation},\ }\href {https://doi.org/10.1143/PTP.63.1217} {\bibfield
  {journal} {\bibinfo  {journal} {Prog. of Theo. Phys.}\ }\textbf {\bibinfo
  {volume} {63}},\ \bibinfo {pages} {1217} (\bibinfo {year}
  {1980})}\BibitemShut {NoStop}%
\bibitem [{\citenamefont {Faraoni}(2015)}]{VF}%
  \BibitemOpen
  \bibfield  {author} {\bibinfo {author} {\bibfnamefont {V.}~\bibnamefont
  {Faraoni}},\ }\bibfield  {title} {\bibinfo {title} {Cosmological and black
  hole apparent horizons},\ }\href {https://doi.org/10.1007/978-3-319-19240-6}
  {\bibfield  {journal} {\bibinfo  {journal} {Lecture Notes in Physics}\
  }\textbf {\bibinfo {volume} {907}},\ \bibinfo {pages} {1} (\bibinfo {year}
  {2015})}\BibitemShut {NoStop}%
\end{thebibliography}%

\end{document}